\documentclass[preprint,9pt]{elsarticle}
\usepackage{multirow}

\usepackage{amssymb}
\usepackage{amsmath}

\usepackage[table]{xcolor}
\usepackage{tikz}
\usepackage{makecell}
\usepackage{graphicx}
\usepackage[font=footnotesize,subrefformat=parens]{subcaption}
\usepackage{amsmath,amssymb,amsfonts}
\usepackage{diagbox}
\usepackage[hyperfootnotes=false]{hyperref}
\usepackage{bm}
\usepackage[symbol]{footmisc}   
\RequirePackage{snapshot}   

\newcommand{\variant}[0]{\hspace{0.1cm}\rotatebox[origin=c]{180}{$\Lsh$}\hspace{0.1cm}}
\newcommand{\subvariant}[0]{\hspace{0.5cm}\rotatebox[origin=c]{180}{$\Lsh$}\hspace{0.1cm}}

\newcommand{\nlt}[0]{\\\noalign{\vskip-0.15pt}}  
\newcommand{\margintxt}[2]{#1~\textbf{(}\textit{#2}\textbf{)}}

\newcommand{\derbg}[1]{\textbf{#1}}
\newcommand{\cder}[1]{(\textit{#1})}    

\newcommand{\vth}[1]{\rotatebox[origin=c]{90}{#1}}
\newcommand{\vthml}[1]{\vth{\parbox{2.4cm}{\centering #1}}}

\DeclareMathOperator*{\argmax}{arg\,max}

\newcommand{\vect}[1]{\mathbf{#1}}

\journal{Computer Speech \& Language}

\begin{document}

\begin{frontmatter}



\title{Dissecting the Segmentation Model of End-to-End Diarization with Vector Clustering}


\author[irit,jsps]{Alexis Plaquet} 
\ead{alexis.plaquet@irit.fr}
\author[ntt]{Naohiro Tawara}
\ead{mail@mail.com}
\author[ntt]{Marc Delcroix}
\ead{mail@mail.com}
\author[ntt]{Shota Horiguchi}
\ead{mail@mail.com}
\author[ntt]{Atsushi Ando}
\ead{mail@mail.com}
\author[ntt]{Shoko Araki}
\ead{mail@mail.com}
\author[irit]{and Herv\'{e} Bredin}
\ead{bredin@pyannote.ai}

\affiliation[irit]{
    organization={IRIT, Université de Toulouse, CNRS, Toulouse INP, UT3},
    addressline={Toulouse}, 
    country={France}
}
\affiliation[ntt]{
    organization={NTT Corporation},
    country={Japan}
}

\fntext[jsps]{This work was done during an internship at NTT as a JSPS International Research Fellow.}

\begin{abstract}
End-to-End Neural Diarization with Vector Clustering is a powerful and practical approach to perform Speaker Diarization. Multiple enhancements have been proposed for the segmentation model of these pipelines, but their synergy had not been thoroughly evaluated. In this work, we provide an in-depth analysis on the impact of major architecture choices on the performance of the pipeline. We investigate different encoders (SincNet, pretrained and finetuned WavLM), different decoders (LSTM, Mamba, and Conformer), different losses (multilabel and multiclass powerset), and different chunk sizes. Through in-depth experiments covering nine datasets, we found that the finetuned WavLM-based encoder always results in the best systems by a wide margin. The LSTM decoder is outclassed by Mamba- and Conformer-based decoders, and while we found Mamba more robust to other architecture choices, it is slightly inferior to our best architecture, which uses a Conformer encoder. We found that multilabel and multiclass powerset losses do not have the same distribution of errors. We confirmed that the multiclass loss helps almost all models attain superior performance, except when finetuning WavLM, in which case, multilabel is the superior choice. We also evaluated the impact of the chunk size on all aforementioned architecture choices and found that newer architectures tend to better handle long chunk sizes, which can greatly improve pipeline performance. Our best system achieved state-of-the-art results on five widely used speaker diarization datasets.
\end{abstract}


\end{frontmatter}

\section{Introduction}

Speaker diarization is a task commonly summarized as ``who spoke when?''. Given an audio file as input, we need to determine at each point in the audio which speakers are active. The task is ``speaker invariant'', meaning we are not concerned with the exact identities of the speakers; we only need to be able to distinguish them.

This task was historically solved with vector clustering-based methods~\cite{reynolds2005diarization}; by combining a voice activity detection system to detect the presence of speech with the extraction of frame-wise speaker embeddings, we can assign a speaker to each active frame using a clustering algorithm. The clustering algorithm takes in a set of vector embeddings and returns a set of clusters. Each of the input embeddings belongs to only one cluster, and all embeddings in one cluster should share similar properties (here belonging to the same speaker). In clustering-based speaker diarization, we assign a cluster to each frame, aiming to get one cluster for each speaker in the audio. Clustering methods have evolved over the years with better embedding extractors (i-vectors~\cite{dehak2011ivectors}, x-vectors~\cite{snyder2018xvectors}), and clustering algorithms (hierarchical agglomerative clustering~\cite{ajmera2003robust,anguera2006robust}, Bayesian HMM~\cite{landini2022vbx}). While they make robust and performant systems that can handle any number of speakers and recording durations, clustering-based approaches cannot handle overlapped speech easily.

With the rise of deep learning approaches, end-to-end neural diarization (EEND) was proposed to solve the issues of the clustering-based approaches~\cite{fujita2019eend}. EEND relies on a single neural network that takes in an input audio and directly outputs the final diarization result. It can handle overlapped speech by design. Each output of the EEND model consists of the estimated speech activity of a speaker in the recording. This implies that the neural network should be able to keep track of speaker activities over long recordings, which brings some challenges. First, the number of outputs should be set to the maximum number of speakers that can be seen in the recordings of the target applications. The network typically needs to be trained with data up to that maximum number of speakers and cannot generalize well to an arbitrary number of speakers. Moreover, EEND has high memory costs (especially during training) induced by the need to process long recordings at once. Because of these limitations, fully EEND systems have been mostly applied to scenarios with relatively short durations or with a limited number of speakers~\cite{fujita2019attention,horiguchi2022eendattractor}.

Since EEND has its own limitations, another approach, ``end-to-end neural diarization with vector clustering'' (EEND-VC), was proposed to bring the best of both clustering-based methods and EEND-based methods~\cite{EEND-vector-clustering_ICASSP2021}. Its principle is to split the input recording into smaller chunks, obtain local segmentations with an EEND model, and then aggregate them into a full file-length diarization through a subsequent vector-clustering step. We call local segmentation the diarization output of an EEND model on a chunk of audio. Since we perform EEND independently for each chunk, the speaker identities are not guaranteed to be consistent across chunks, i.e., the $i$-th activities of two local segmentations do not necessarily belong to the same speaker. The vector-clustering step is needed to solve the permutation ambiguity problem that would otherwise prevent aggregating local segmentations. This two-step process allows the processing of arbitrarily long recordings (by using the EEND model with small chunks) with arbitrarily high numbers of distinct speakers (by relying on clustering algorithms) while keeping the advantage of EEND of handling overlapped speech.

EEND-VC provides a practical approach to the diarization problem, consequently, there have been many works following this framework \cite{bredin23pyannote21,baroudi2023pyannote,delcroix23_interspeech,tawara2024chime7,han2024leveragingselfsupervisedlearningspeaker}. The EEND-VC framework can be implemented in multiple ways, while in the original framework the EEND model predicts both the speaker activity and the speaker embeddings for each chunk, the pyannote framework instead combines an EEND model which only predicts speaker activity with pre-trained speaker embedding extractor to achieve EEND-VC~\cite{bredin23pyannote21}.
Since the pyannote framework relies on a generic EEND model, it can directly rely on recent advances in EEND architectures. Different encoders have been used, such as BLSTM~\cite{fujita2019eend,bredin23pyannote21}, Transformer~\cite{fujita2019attention,EEND-vector-clustering_ICASSP2021}, Conformer~\cite{han2024leveragingselfsupervisedlearningspeaker}, etc. Features extraction has been done with log-mel filterbanks~\cite{fujita2019eend}, trainable SincNet filterbanks~\cite{ravanelli2018sincnet,bredin23pyannote21} or pretrained WavLM~\cite{wavlm,baroudi2023pyannote}. Although many system configurations have been proposed, there is a lack of a fair comparison between the previously proposed systems as they usually differ in terms of architectures, training, and evaluation data.

In this paper, we investigate the architecture choices of EEND-VC models. First, we introduce two modifications to the local EEND model of an EEND-VC system, which are based on our previous conference papers~\cite{plaquet23powerset,plaquet2024mambadiarization}. The first modification concerns the model architecture, where we introduce a module based on Mamba \cite{gu2024mamba,plaquet2024mambadiarization}. Mamba is a kind of state-space model that has been shown to be powerful for modeling long sequences and was used for several speech tasks such as speech enhancement, speech recognition, and speech synthesis \cite{zhang2024mamba,miyazaki24_interspeech}. However, it had not been used for speaker diarization, despite the fact that diarization handles speech sequences typically longer than most other speech tasks.

The second modification concerns the training objective of the model. Historically, the speaker diarization task is framed as a multilabel problem using the binary cross entropy loss~\cite{fujita2019eend}. Instead, we introduced a powerset representation for the speaker activity, which allows for reframing the multilabel problem into a multiclass problem optimized with a cross-entropy loss~\cite{plaquet23powerset}. We expect that the multiclass optimization problem simplifies the training of the diarization system, allowing better performance on complex architectures. 

Another important factor for EEND-VC is the chunking duration (also called chunk size), which has not been sufficiently investigated in past studies \cite{tawara2024chime7}.
With very short chunks (e.g., 1s), the EEND model does not need to track speakers for a long duration and acts similarly to a voice activity detection (VAD) model. However, it becomes hard to estimate reliable speaker embeddings, especially when speech overlaps in a chunk. The pipeline ends up thus very similar to more classical clustering pipelines, with poor handling of overlapped speech.
On the other hand, with long chunks (e.g., 120s), the EEND model needs to track speakers over a long duration, which may be challenging for some model architectures. Nonetheless, if the model can provide correct segmentations, the VC part becomes simpler because Speaker embeddings can be extracted on many more audio samples, making them more accurate and easier to cluster. 
We can see that there must be a tradeoff between these two extremes. Since existing work on varying chunk lengths in the EEND-VC pipeline is sparse \cite{tawara2024chime7}, we aim to clarify experimentally this tradeoff for various model configurations and datasets.

Our investigation leads us to an extensive evaluation covering 120 different configurations of the EEND-VC framework under the same evaluation protocol. In particular, we compare different encoders (SincNet vs. WavLM), decoders (LSTM, Conformer, Mamba), and training loss (multilabel vs. powerset multiclass). For each investigated factor, we look at its impact on the quality of the local segmentation and final diarization. Besides, we also investigate the impact of the chunk size on the performance of the different models.

This paper is based on our prior works using powerset \cite{plaquet23powerset} and Mamba for EEND-VC \cite{plaquet2024mambadiarization} and includes more discussions and experimental analyses. 
It extends \cite{plaquet23powerset}, which showcased the advantages of the powerset loss but only did so on one EEND architecture with one chunk size. Here, we study its performance on a wide variety of EEND-VC model configurations and chunk sizes. It also extends \cite{plaquet2024mambadiarization}, which showcased the performance gained by using the Mamba block as the decoder but lacked comparison against a solid attention-based processing-module, and only did so with a WavLM Base encoder. Here, we include evaluations with systems using SincNet, a lightweight feature extraction, and comparisons with a strong Conformer model.

The paper's outcome is an extensive evaluation of various configurations of EEND-VC frameworks under the same evaluation protocol, which allows for a better understanding of the implications of architectural choices for different design requirements. In particular, we find that chunk size is a crucial pipeline hyperparameter. It can highly improve performance if the EEND architecture is powerful enough to deal with long sequences, which is the case with Mamba and Conformer. We obtain comparable results when using Mamba or Conformer, but Mamba's performance is more stable across different configurations, while Conformer achieves slightly better optimal performance under the optimal configuration. Similarly, the powerset loss seem to facilitate training, but compared to multilabel, it hinders performance for powerful enough models. This notably happens if we unfreeze WavLM weights to allow finetuning, which is crucial to attain state-of-the-art (SOTA) results. Through these experiments, we achieve new SOTA on five out of the eight datasets we investigate.

In the remainder of this paper, we provide an overview of the EEND-VC framework in \autoref{sec:eendvc-overview}, and detail the different configurations for the EEND model in \autoref{sec:segmentation-model}, including our proposed Mamba-model and multi-class powerset loss. \autoref{sec:experimental-protocol} describes the experimental protocol and Section \autoref{sec:results} discusses the results. Finally, \autoref{sec:discussion} summarizes the experimental findings, and \autoref{sec:conclusion} concludes our paper.

\section{Overview of EEND-VC framework}
\label{sec:eendvc-overview}

\begin{figure}[tb]
    \centering
    \includegraphics[width=\linewidth]{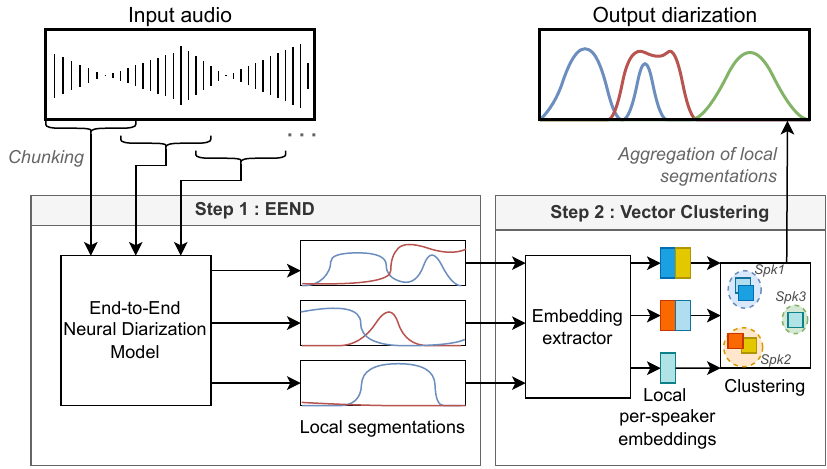}
    \caption{Overview of an EEND-VC pipeline.}
    \label{fig:eendvc_pipeline}
\end{figure}

This study is focused on the EEND-VC framework, which consists of two distinct steps, an ``EEND'' step and a ``Vector Clustering'' step, as shown in \autoref{fig:eendvc_pipeline}. We summarize these steps below.

The first step is the ``EEND'' part of the EEND-VC process, which relies on a neural network to estimate the activity of the speakers in the input speech signal. Here, we assume that the EEND model can handle up to $N$ speakers. This limit is present in current architectures, either due to a limitation of the architecture or a limitation of the training data. The EEND model directly provides the speech activity of $N$ speakers in the input audio waveform. However, current SOTA EEND models cannot process arbitrarily long files as they must keep the whole sequence in memory due to their architecture, which is usually based on attention or bidirectional RNNs.

To remedy these problems, we cut the input waveform $\vect{x}$ (possibly hours long) into $L$ shorter audio chunks of duration $W$, composed of $W_x$ frames,
\begin{align}
    [\vect{x}_1, \dots, \vect{x}_L] &= \text{chunking}(\vect{x}) , 
\end{align}
where $\vect{x}_l \in \mathbb{R}^{W_x}$ is the speech signal associated with the $l$-th chunk.
These small chunks are then fed to the EEND model to obtain ``local segmentations'' as,
\begin{align}
    \tilde{\vect{y}_l} &= f_{\text{EEND}}(\vect{x}_l) , 
\end{align}
where $\tilde{\vect{y}_l}\in \mathbb{R}^{W_y \times N}$ consists of the speaker activity posteriors for all $N$ speakers in the chunk, and $W_y$ is the number of frames at the output of the EEND model.
After obtaining the segmentations for each chunk, we need to aggregate them to obtain diarization results for the whole recording. This is not trivial because of the speaker permutation ambiguity problem.

The second step of the pipeline is the ``Vector Clustering'' part, which resolves the speaker permutation ambiguities by realigning all speaker identities in all local segmentations.
Vector clustering first extracts speaker embeddings for each speaker in each local segmentation (extracting at most $N$ speaker embeddings per local segmentation) as,
\begin{align}
    \mathbf{e}_{l} = \text{Embedding}(\vect{x}_l, \vect{y}_l),
\end{align}
where $\mathbf{e}_{l} \in \mathbb{R}^{D \times N}$ are the $N$ speaker embeddings for the $l$-th chunk, and $\text{Embedding}(\cdot)$ is a speaker embedding extraction neural network such as ECAPA-TDNN \cite{desplanques20ecapatdnn}.
We then perform clustering on the embeddings to find the $N'$ global identities (i.e., the clusters) as
\begin{align}
     [\phi_1, \dots, \phi_L] &=\text{vector clustering}([\vect{e}_1, \dots,\vect{e}_L]),
\end{align}
where $\text{vector clustering}(\cdot)$ is a function representing the clustering process. It outputs a mapping $\phi_l: \{1,\dots,N\} \rightarrow \{1,\dots,N'\}$ for each chunk, between the $N$ local speakers and $N'$ cluster centroids that should represent global speaker identities.
Finally, we obtain the final diarization $\tilde{\vect{Y}}\in \mathbb{R}^{L \cdot W_y \times N'}$ by aggregating (concatenating) the local segmentation of all chunks after solving the speaker permutations using the mapping $\phi_l$, and discretizing the resulting aggregation. Discretization is required to obtain clear-cut speaker boundaries instead of probability distributions, and it is done by thresholding the speaker activities ($\in [0,1]$). The default threshold is $0.5$, but it can be treated as a pipeline hyperparameter, as the optimal threshold might change slightly depending on the dataset.

We explained the process when the chunking does not have overlap (i.e., chunk length = step size). Changing the step size so that chunks overlap only affects the reconstruction of the global diarization, $\tilde{\vect{Y}}$. Instead of concatenating the output, we obtain each frame of the output $\tilde{\vect{Y}}_t$ by averaging the speaker activities of the chunks that are temporally aligned with it. In practice, overlapping chunks are preferred; they result in more accurate prediction of speaker activities, as the final output frames are obtained from the average of multiple predictions that each have access to a slightly different context (earlier in the audio, or later in the audio).

\section{EEND model for local segmentation}
\label{sec:segmentation-model}

\begin{figure}[tb]
    \centering
    \includegraphics[width=\linewidth]{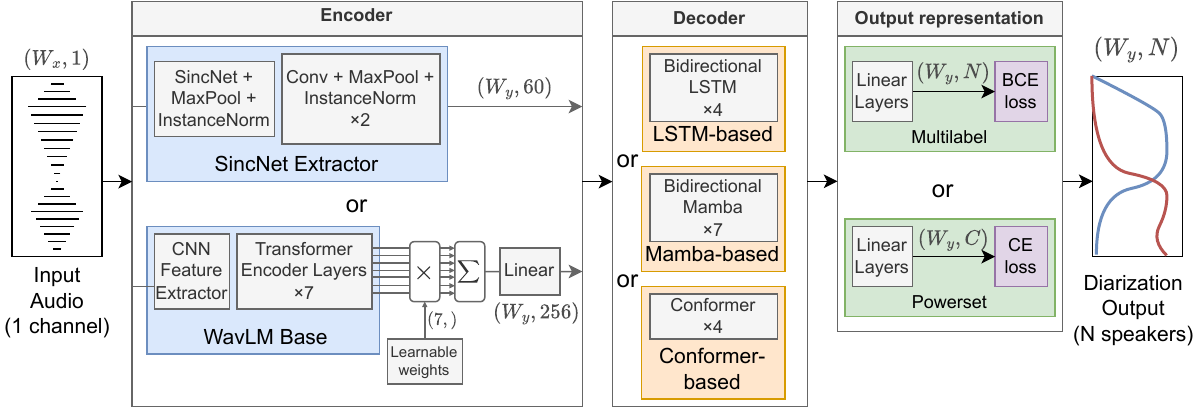}
    \caption{Proposed architectures for our EEND segmentation models. We test each possible combination of encoder, decoder, and training loss.}
    \label{fig:eend_architecture}
\end{figure}
In this paper, we are interested in the impact on speaker diarization of different architectures of the EEND model used to obtain the local segmentation for EEND-VC. The different architectures we investigate in this paper are summarized in \autoref{fig:eend_architecture}. We can divide EEND models into three main parts: the encoder, the decoder, and the training loss. 

The encoder downsamples the signal and extracts features that are then processed by a decoder, followed by a few linear layers and an activation function that will determine the training loss used. We evaluate two encoders (SincNet and WavLM), three decoders (LSTM, Conformer, and Mamba), and two training losses (multi-label and multi-class with powerset). We explain each of them in this section with more details for the Mamba-based decoder and the multi-class powerset loss, which are based on our prior works. All possible combinations of these modules are investigated to determine the impact of each factor.

\subsection{Encoder}

The encoder is responsible for downsampling the input waveform from a few kHz to a feature sequence with a lower time resolution (usually around 50Hz), capturing meaningful information at each frame. We study two encoders: SincNet which is lightweight, and WavLM which provides better speech representations at the cost of significantly more parameters and layers.

\subsubsection{SincNet}

SincNet \cite{ravanelli2018sincnet} is a type of learnable filterbank, which is an alternative to fixed filterbanks (such as Short-Term Fourier Transform or Mel Spectrogram filterbanks) and learnable convolution filterbanks. With SincNet, each filter is composed of two parameters: the cutoff frequencies $f_1$ and $f_2$, which are learnable through backpropagation. SincNet is named after the $sinc$ function, which is obtained when converting the $rect$ function (used for bandpass filters) from the frequency to the temporal domain.

SincNet is light in memory, fast, and conceptually simple. In speaker diarization, its use is preferred over fixed filterbanks, which cannot be optimized for the task, and naive convolution-based filterbanks, which are much heavier in parameter count and harder to train~\cite{bredin2019pyannotebuildingblocks, bullock2020overlapawarediar}. Despite being easier to train than most other convolution-based filterbanks, SincNet still requires a large quantity of data and training steps to converge.

The pyannote toolkit \cite{bredin23pyannote21, bredin2019pyannotebuildingblocks} on which we base our experiments uses a SincNet-based encoder as the default encoder of its default EEND model. It is composed of a SincNet layer and two convolutions, all of them are followed by a max pooling and instance normalization operations.

\subsubsection{WavLM}

WavLM is a large attention-based self-supervised learning (SSL) model trained with a pretext task on noisy overlapped speech, with publicly available code and checkpoints \cite{wavlm}. Pretrained weights are available for two architectures: WavLM Base, which outputs 768 features per frame for each of its 12 encoder layers, and WavLM Large, which outputs 1024 features per frame for each of its 24 encoder layers. WavLM enabled multiple state-of-the-art results in speaker diarization \cite{wavlm,baroudi2023pyannote, delcroix23_interspeech,han2024leveragingselfsupervisedlearningspeaker} and WavLM Base outperforms similar SSL-based encoders~\cite{baroudi24_interspeech} such as HuBERT~\cite{hsu2021hubert} or Wav2Vec2.0~\cite{baevski2020wav2vec2}. We focus on the WavLM Base architecture, as it is three times smaller than the large variant and has already been used in multiple prior works with great success. Following the methodology in the WavLM paper, proposed by the SUPERB benchmark \cite{yang21csuperb}, we extract features from the pretrained model with a learnable weighted sum of the features obtained after each of its layers. We reduce the number of features from 768 to 256 through a Linear layer, as we found in \cite{plaquet2024mambadiarization} that it greatly increases convergence rate on preliminary experiments, this choice is supported by \cite{han2024leveragingselfsupervisedlearningspeaker} which reduces to the same number of features.

WavLM consists of a stack of convolutional layers followed by multiple Transformer blocks with self-attention.
Compared with SincNet, which comprises only a set of convolutional layers, WavLM has much more processing power.
When looking at performance gaps between the encoders in this paper, we need to keep in mind that WavLM does a lot more than a simple filterbank followed by convolutions.
It utilizes self-attention to access the entire input sequence, allowing it to consider the whole context of the sequence to generate each frame-wise representation. In contrast, SincNet’s receptive field is limited to just 62 ms.

\subsection{Decoder}

\begin{figure}
    \centering
    \begin{subfigure}[b]{0.22\textwidth}
         \centering
         \includegraphics[width=\textwidth]{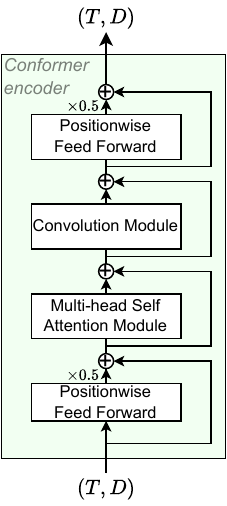}
         \caption{Conformer encoder block.}
         \label{fig:encoder_architectures_conformer}
     \end{subfigure}
     \hfill
     \begin{subfigure}[b]{0.74\textwidth}
         \centering
         \includegraphics[width=\textwidth]{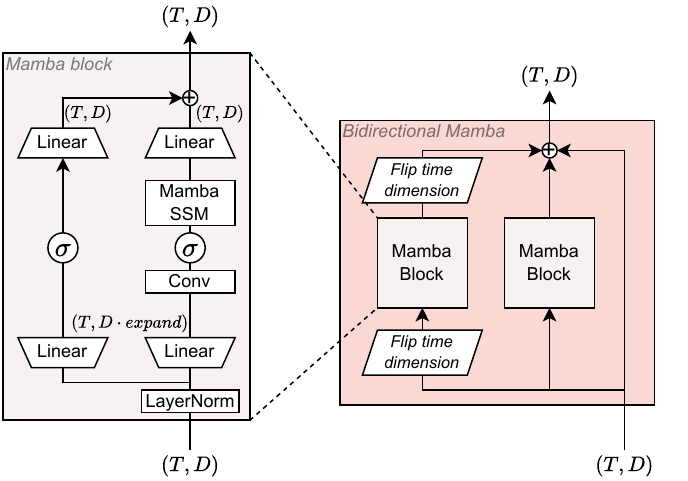}
         \caption{Mamba block (left) and the bidirectional Mamba block (right).}
         \label{fig:encoder_architectures_mamba}
     \end{subfigure}
    \caption{Overview of (a) the Conformer block and (b) the (Bidirectional) Mamba block. $+$ is the elementwise addition, $\sigma$ is the Swish activation function.}
    \label{fig:encoder_architectures}
\end{figure}

The decoder is what we call the blocks that will solve the diarization task from input audio features. It is usually implemented by stacking several layers or blocks, such as LSTMs, Conformer, or Mamba.

\subsubsection{LSTM}

Long short-term memory (LSTM) networks are a type of recurrent neural networks (RNNs)~\cite{hochreiter1997lstm}. They take a sequence of inputs (in our case, a sequence of audio features) and, for each step in the sequence, give a prediction while updating an internal state that serves as memory. This memorization mechanism is crucial for diarization, as this is how the module stores the identities of previously seen speakers so that the model can predict consistent speaker activities through its output.
If memorization fails, we might encounter speaker confusion errors.

While LSTMs were introduced in 1997, simple Bidirectional LSTMs (BiLSTM) have managed to remain competitive in the field of speaker diarization until recently \cite{plaquet23powerset}. This is in large part thanks to the EEND-VC approach, which allows working on short chunks (which contain fewer speakers and require less memorization of speaker identities) and relies on VC to compensate for LSTMs' weak memorization capabilities.

\subsubsection{Conformer}

The conformer decoder block~\cite{gulati20conformer} is based on the transformer decoder block~\cite{vaswany2017transformer}. It achieves better results on audio data by integrating convolutional neural networks (CNN) into the architecture in order to better capture local features. An overview of the Conformer decoder architecture is presented in \autoref{fig:encoder_architectures_conformer}. Like the transformer block, it is based on multi-head self-attention, which considers the whole sequence at once, unlike RNNs. This effectively removes the need to design a memorization mechanism, at the cost of much increased memory requirements. It also tends to be faster on long sequences since RNNs have to step through every element of the sequence, while self-attention can fully parallelize this processing.

The Conformer block was proposed to improve the performance of automatic speech recognition systems and has been used to achieve SOTA results in speaker diarization \cite{han2024leveragingselfsupervisedlearningspeaker,yang2024nsdms2s}. Interestingly, EEND papers that rely on attention-based blocks find that positional embedding does not help the EEND model \cite{fujita2019attention}. It essentially means they act as clustering algorithms on the extracted audio features, completely disregarding data sequentiality.

\subsection{Mamba}

In the last few years, research on state space models (SSMs) for deep learning has resulted in increasingly performant architectures. Among them, the recently proposed Mamba SSM block \cite{guEfficientlyModelingLong2022} reached performance on par with self-attention. 

Recently, several works have explored the use of Mamba for speech applications, including speech enhancement, speech recognition, text-to-speech, spoken language understanding, and speech summarization \cite{zhang2024mamba, miyazaki24_interspeech}. 
These studies found that Mamba achieves comparable or slightly better results than Transformer-based architecture and does well on long sequences \cite{miyazaki24_interspeech}. These prior studies motivated us to investigate the potential of Mamba for speaker diarization.

\subsubsection{Overview of SSM and Mamba blocks}

An SSM is a mathematical framework to describe the evolution of a model in continuous time, while they have been applied across many fields of research, usage of SSM in deep learning is still fairly recent \cite{gu2021lssl}. This model possesses inputs, outputs, and an internal state. Such a system can be described by a set of two equations,
\begin{align}
    \vect{h}'(t) & = \vect{A}(t)\vect{h}(t) + \vect{B}(t)\vect{u}(t), \label{eq:ssm1} \\
    \vect{z}(t) & = \vect{C}(t)\vect{h}(t) + \vect{D}(t)\vect{u}(t),  \label{eq:ssm2}
\end{align}
where $\vect{u}(t)$, $\vect{h}(t)$ and $\vect{z}(t)$ are the input, state and output vectors at time $t$, respectively. $\vect{A}(t)$, $\vect{B}(t)$, $\vect{C}(t)$, $\vect{D}(t)$ are state, input, feedforward, and output matrices, respectively.
\autoref{eq:ssm1} defines the evolution of the state vector $\vect{x}(t)$ over time through a differential equation. This derivative depends on the state vector $\vect{x}$, and the input vector $\vect{u}$.
\autoref{eq:ssm2} defines the output vector $\vect{z}$ over time and depends on the state $\vect{h}$, and the input vector $\vect{u}$. In order to allow a description of a non-linear system, all vectors and matrices in the system are time-dependent (i.e., depend on $t \in \mathbb{R}$).

To use an SSM as a general deep learning block, we need a way to obtain matrices $\vect{A}$, $\vect{B}$, $\vect{C}$, and $\vect{D}$ for each frame of the input. 
However, this definition of SSM is continuous, while machine learning mostly deals with discrete inputs and output sequences (whether it is a sequence of tokens or a discrete audio waveform). Fortunately, the two equations can be discretized, which adds another parameter $\Delta$ (the time between two steps of the system) as described in \cite{gu2024mamba}. For example, the discretized version of $\vect{A}$ can be obtained as $\bar{\vect{A}} = \text{exp}(\Delta\vect{A})$.

Previous SSM-based architectures such as \cite{guEfficientlyModelingLong2022} simplified the model by relying on time-invariant matrices (i.e. $\vect{A}$, $\vect{B}$, $\vect{C}$, and $\vect{D}$ do not depend on $t$), this makes the model less powerful (linear) but allows very fast execution. However, Mamba's SSM parameters \textit{are} time-dependent: $\vect{B}(t)$, $\vect{C}(t)$, $\vect{D}(t)$ and $\vect{\Delta}_t$ are predicted at each frame through Linear layers. $\vect{A}$ is time invariant, but making $\vect{\Delta}_t$ time-dependent is akin to making its discretized version $\bar{\vect{A}}(t)=\exp(\Delta_{t}\vect{A})$ time-dependent. Mamba's SSM is also called a ``selective SSM'' due to its ability to select whether the input should be saved into the state depending on its content.

After discretization, the selective SSM behaves as a recurrent neural network and processes sequentially the input sequence while updating its internal state. The gating mechanisms of this SSM can be considered as a generalization of those found in other existing RNN architectures. Thanks to its optimized implementation, Mamba has simultaneously low memory usage, fast inference, and attention-like processing capabilities.

\subsubsection{Mamba for speaker diarization}
Our implementation of the Mamba-based EEND model simply replaces the LSTM blocks with a stack of Mamba blocks. 
The original Mamba paper~\cite{gu2024mamba} introduced a Mamba block with the selective SSM as its core, which was designed to maximize its performance. The Mamba block is represented in \autoref{fig:encoder_architectures_mamba}. It first performs layer normalization and then has two parallel processing paths, a stack of two linear layers, and the Mamba SSM block sandwiched with linear and convolutional layers.
In this figure, $expand$ is a hyper-parameter that defines the factor by which the number of features is multiplied inside the block. The selective SSM itself has multiple parameters, such as $d\_state$ the size of $\vect{h}$, and the number of output features of the Linear layer used to predict $\Delta_{t}$.

We employ a bidirectional Mamba block similar to that proposed in \cite{zhang2024mamba} for speech enhancement and speech recognition. There are two ways of implementing bidirectional Mamba: by duplicating the SSM inside the Mamba block (dubbed ``Inner Bidirectional Mamba'') or by duplicating the Mamba block (dubbed ``External Bidirectional Mamba''). \cite{zhang2024mamba} found that the latter obtains superior performance over the former and is the one we choose and call Bidirectional Mamba (or BiMamba) in this paper. 
The bidirectional Mamba block simply consists of two parallel Mamba blocks, with one receiving the feature sequence flipped in time. The outputs of each Mamba block are added together with the input to realize a skip connection.

\subsection{Training loss}

Finally, we discuss two options for the training loss of the EEND model.

\subsubsection{Multilabel with binary cross entropy loss}

The original framework proposed for EEND formulates it as a multilabel classification problem, each label $y_i \in [0,1]$ is the activity of speaker $S_i$ (where $\{S_1,\dots,S_N\}$ are the $N$ speakers) \cite{fujita2019eend}. For example, the reference vector $\vect{y} = [y_1,\dots,y_N]^{\intercal}$ is defined as
\begin{equation}
    y_i = 
    \begin{cases}
        1 & \text{if the speaker}~S_i~\text{is active,} \\
        0 & \text{otherwise,}
    \end{cases}
\end{equation}
where we omit the temporal dimension for clarity.
In this framework, the model takes a chunk of audio $\vect{x}$ as input and produces an output $\tilde{\vect{y}} \in [0,1]^{W_y\times N}$ (where we omitted the chunk index for simplicity). The last layer after the decoder consists thus of a linear projection layer, which maps the intermediate representation to $N$ outputs. It is followed by a sigmoid activation function that constrains the outputs of a network in $[0,1]$, to obtain speech activity posteriors for each speaker. These posteriors can be thresholded to obtain a discrete diarization prediction with clear beginnings and ends of speaker activity.

Speaker diarization is permutation-invariant to speaker identities, i.e., it only matters to allocate consistent speaker labels within a recording and not assign absolute speaker labels.
Therefore, to compute the training loss $L$ from the loss function $\mathcal{L}$, reference $\vect{y}$ and prediction $\tilde{\vect{y}}$, we need to find the best possible permutation of speakers $\pi^*$ as,
\begin{equation}
    L = \min_{\pi} \mathcal{L}( \pi (\tilde{\vect{y}}) , \vect{y}),
    \label{eq:loss_perm}
\end{equation}
where $\mathcal{L}$ is the training loss, which here consists of the binary cross entropy (BCE).

\subsubsection{Multiclass powerset with cross-entropy loss}

\begin{figure}[tb]
    \centering
    \includegraphics[width=1\linewidth]{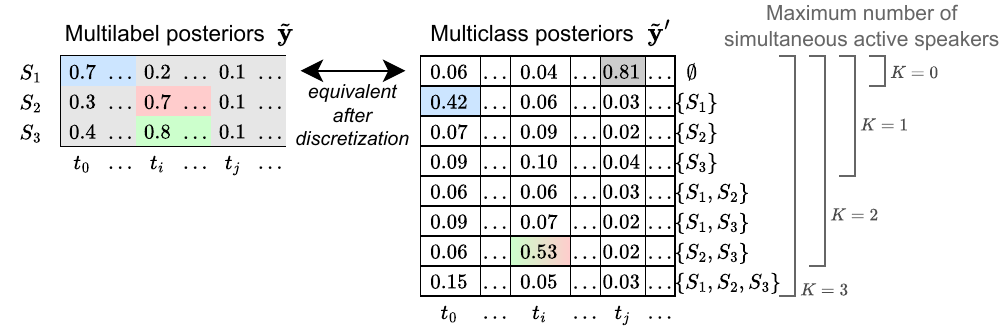}
    \caption{Comparison of multilabel representation and multiclass powerset representation for $N=3$. Active speakers in the multilabel representation are obtained with thresholding ($>0.5$), and active speakers in the multiclass are obtained from the argmax class.}
    \label{fig:powerset}
\end{figure}

\begin{figure}[tb]
    \centering
    \includegraphics[width=0.4\linewidth]{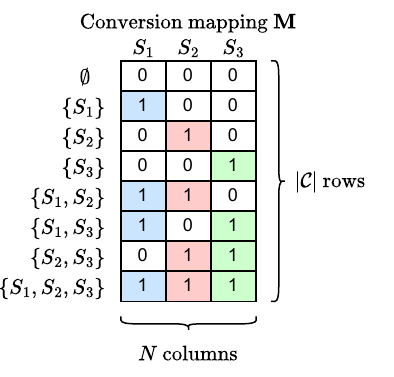}
    \caption{Example of a powerset $\leftrightarrow$ multilabel mapping with $N=3$.}
    \label{fig:powerset_mapping}
\end{figure}

Recently, we proposed a multiclass powerset formulation of the EEND speaker diarization task \cite{plaquet23powerset} as an alternative to the multilabel one. 

In the speaker diarization task, multiple speakers can be active on the same frame. To allow this in the multiclass representation, we use the \textit{powerset} encoding: each combination of the $N$ speakers is encoded as one class (e.g., $\emptyset, \{S_1\}, \{S_2\}, \{S_1,S_2\}$, etc.). We show how the multilabel and multiclass powerset representations relate in \autoref{fig:powerset}. 

A multiclass classifier outputs the probability distribution $\tilde{\vect{y}}' = [\tilde{y}'_1, \dots, \tilde{y}'_C]$ of $C$ classes such that $\sum_{c=1}^C \tilde{y}'_c = 1$, and the class of maximum probability $\argmax_{c\in\left\{1,\dots,C\right\}}{\tilde{y}'_c}$ is the predicted class. Unlike with multilabel, an additional thresholding hyperparameter is not needed for the discretization.
When using such a powerset representation, the last layer after the decoder consists of a linear layer that maps the internal representation to $C$ outputs, followed by a softmax layer to obtain state probability distribution.

Using the powerset leads to an exponential scaling of the number of classes $C$ with the number of speakers $N$. However, high numbers of simultaneously active speakers (three or more speakers active simultaneously) are both rare and hard to annotate, which means the exponential scaling can be prevented by limiting the maximum number of speakers in the combination. We call $K$ the maximum number of active speakers encoded, the size of the set of speaker classes $\mathcal{C}_{N,K}$ in the multiclass space for a given $K$ is then given by 
\begin{equation}
    \label{eq:powersetK}
    |\mathcal{C}_{N,K}| = \sum_{k=0}^{K} \tbinom{N}{k}.
\end{equation}
Using \autoref{eq:powersetK}, we can show that using $K=2$ instead of $K=N$, we save one class when $N=3$; however, when $N=6$, we save 42 classes, and when $N=7$, we save 99 classes, and so on.

The default training loss $\mathcal{L}$ for multiclass problems is the cross-entropy loss.
As for the multilabel loss, we need to solve the permutation ambiguity between the prediction and references.
We resolve it by finding the best permutation in the multilabel space, as in \autoref{eq:loss_perm} (with BCE). To do so, we convert the powerset prediction $\tilde{\vect{y}}'$ to the multilabel prediction $\tilde{\vect{y}}$, find the best permutation of the reference $\pi^*(y)$ and convert the permutated reference to the powerset space. To perform such a conversion, we create a conversion mapping matrix $\vect{M} \in (0,1)^{|\mathcal{C}_{N,K}| \times N}$ matrix, defined as
\begin{equation}
    \label{eq:powerset_mapping}
    {M}_{i,j} = 
    \begin{cases}
        1 & \text{if } j \in \mathcal{C}_i \\
        0 & \text{otherwise.}
    \end{cases}
\end{equation}

An example of such mapping matrix with $N=3$ speakers is given in \autoref{fig:powerset_mapping}. To convert a vector $\vect{\tilde{y}}'$ from powerset space to multilabel space, we multiply it by the matrix as in \autoref{eq:ps_to_ml}. This conversion keeps soft probabilities (e.g. $\vect{\tilde{y}}' = [0.9, 0.04, 0.6]^\intercal$ will be converted to the equivalent $\vect{\tilde{y}}$ with soft probabilities).
\begin{equation}
    \label{eq:ps_to_ml}
    \tilde{\vect{y}} =  \tilde{\vect{y}}'^\intercal  \times \vect{M}
\end{equation}

 Conversely, to convert a vector $\vect{\tilde{y}}$ from multilabel to powerset space, we multiply it by the transpose of the matrix and use the argmax as the index of the active class, as shown in \autoref{eq:ml_to_ps}. With this encoding, the argmax function needs to return the lowest index in case of a tie. For example, with $N=3$, if $\vect{\tilde{y}} = [1,0,0]^\intercal$ (speaker $S_1$ active alone), then $\tilde{\vect{y}}^\intercal  \times \vect{M}^\intercal = [0,1,0,0,1,1,0,1]$, which means the maximum value (1) is shared by all classes containing $S_1$ (i.e. $\{S_1\},\{S_1,S_2\},\{S_1,S_3\},\{S_1,S_2,S_3\}$). The correct class is obtained by selecting the class with maximum value and minimal cardinality (i.e. argmax), here $\{S_1\}$. Since the conversion relies on argmax, it does not keep soft probabilities.

\begin{equation}
    \label{eq:ml_to_ps}
    \tilde{\vect{y}}_i' = \begin{cases}1&\text{if } i = \text{argmax}(\tilde{\vect{y}}^\intercal  \times \vect{M}^\intercal )\\0&\text{otherwise.}\end{cases}
\end{equation}

Networks with a multiclass encoding were found to give better results than the multilabel encoding at no additional computational cost \cite{plaquet23powerset}. We found in our previous paper \cite{plaquet2024mambadiarization} that this might be due to an improved convergence rate: models trained with powerset need fewer epochs to reach the same performance as their multilabel counterparts.

\section{Experimental protocol}
\label{sec:experimental-protocol}

We designed an experimental protocol that allows an extensive comparison of the different configurations of the EEND model used in the EEND-VC framework.
The protocol allows comparison over various datasets. In this section, we briefly explain the data we use and the different settings of our EEND-VC pipeline.

\subsection{Data}

\begin{table*}[tb]
    \centering
    \caption{Datasets used in this paper and their characteristics. Speaker count statistics are per-recording.}
    \resizebox{\linewidth}{!}{
    \begin{tabular}{|c|c|c|c|c|c|c|c|}
        \hline
        \multirowcell{2}{Dataset} & \multicolumn{3}{c|}{Length} & \multirowcell{2}{Language} & \multirowcell{2}{Content} & \multicolumn{2}{c|}{Speakers} \\
        \cline{2-4} \cline{7-8} &  Train & Dev & Test & & & avg & max \\
        \hline
        AISHELL-4 & 102h & 2h & 13h & Mandarin & Meetings & 5.8 & 7 \\
        AliMeeting & 111h & 4h & 11h & Mandarin & Meetings & 3.0 & 4 \\
        AMI & 80h & 10h & 9h & English & Meetings & 3.9 & 4 \\
        MSDWild \textit{Few} & 64h & 2h & 14h & Multi & In-the-wild & 2.6 & 4 \\
        NOTSOFAR-1 & 30h & 10h & 16h & English & Meetings & 4.7 & 7 \\
        MagicData-RAMC & 150h & 10h & 21h & Mandarin & One-on-one call & 2.0 & 2 \\
        VoxConverse & 18h & 2h & 43h & English & In-the-wild & 6.5 & 21 \\
        Synthetic & 1265h & 7h & ----- & English & 4-speakers Synthetic & 4.0 & 4 \\
        \hline
        DIHARD-3 & ----- & ----- & 33h & English & Various & 2.9 & 9 \\
        \hline
    \end{tabular}
    }
    \label{tab:datasets}
\end{table*}

We trained all models using a compound set consisting of eight datasets: AISHELL-4~\cite{AISHELL}, AliMeeting~(SDM) \cite{AliMeeting}, AMI (SDM)~\cite{AMI}, MSDWILD (Few)~\cite{msdwild}, NOTSOFAR-1~\cite{notsofar1}, MagicData-RAMC~\cite{RAMC}, VoxConverse~\cite{VoxConverse}, and a synthetic dataset. For the NOTSOFAR-1 dataset, we used Train Sets 1 and 2 as the training set, and Train Set 3 as the development set, as defined in the first period of CHiME-8 challenge\footnote{\url{https://www.chimechallenge.org/challenges/chime8/task2/data}}. For single-distant microphone (SDM) datasets, we only use the first channel of the microphone array. All audio files are sampled at 16kHz. The synthetic dataset consists of 50-second simulated multi-talker recordings of up to four speakers. It was created following \cite{yamashita2022improvingnaturalnesssimulatedconversations}, using speech signals from LibriSpeech \cite{librispeech}, noise signals from MUSAN \cite{musan} and room impulse responses \cite{ko2017reverberant}. The characteristics of each dataset are summarized in \autoref{tab:datasets}. 

For evaluation, we did not evaluate the synthetic data (as it is not a standard dataset). We also included the DIHARD-3 dataset for final evaluations (but not during training) to evaluate how well models handle domain shifts.

\subsection{Settings}

\subsubsection{EEND-VC pipeline}

\begin{table}[tb]
    \centering
    \caption{Hyper-parameters used during hierarchical agglomerative clustering for each chunk size.}
    \begin{tabular}{|c|c|c|}
        \hline
        \makecell{Chunk\\size (s)} & \makecell{Clustering\\threshold} & \makecell{Minimum\\cluster size} \\
        \hline
        5 & 0.6915 & 10 \\
        10 & 0.6836 & 7 \\
        30 & 0.6791 & 6 \\
        50 & 0.6846 & 6 \\
        \hline
    \end{tabular}
    \label{tab:clust_hparams}
\end{table}

We implemented our EEND-VC systems following the pyannote 2.1 pipeline \cite{bredin23pyannote21}. For its EEND model, the pipeline used SincNet-based encoder, an LSTM-based decoder, multilabel representation, and a chunk size of $W=5$ seconds.

We only changed the EEND model for our proposed EEND model architectures. Our \textit{(SincNet+LSTM)}-based models are almost identical replicas of the models found in previous pyannote-based papers \cite{bredin23pyannote21,plaquet23powerset}\footnote{Note that our training scheme and our compound set differ from pyannote's default ones. However, we confirmed that the results of our baseline (SincNet+LSTM) were equivalent to pyannote's default configuration.}. For all architectures investigated, we trained four variants of chunk size $W \in \{5, 10, 30, 50\}$ seconds. In all cases, we set the step size to $0.2 \cdot W$. 

For the vector clustering, we used a ResNet model \cite{wang2023wespeaker} trained on VoxCeleb \cite{VoxCeleb17,VoxCeleb18,VoxCeleb19} to extract chunk-wise speaker embeddings. Embeddings were extracted only on non-overlapped speech. The clustering of chunk-wise speaker features was done using hierarchical agglomerative clustering with centroid linkage, following the pyannote EEND-VC pipeline~\cite{bredin23pyannote21}.

The agglomerative clustering contains two hyper-parameters: the clustering threshold and the minimum cluster size, which control the number of clusters (speaker identities) of the predictions. We share hyper-parameters values when the chunk size is identical, the tuned values we used are shown in \autoref{tab:clust_hparams}. In the best-case scenario (enough computational resources, enough validation data), they should be optimized after everything else to bring the best possible performance. However, we observed that chunk duration and step influence the clustering hyper-parameters the most. Since our paper contains more than 200 experiments in total, we decided to share hyper-parameters for all models of equal chunk size in order to make the paper feasible with our resources. To find the shared clustering hyper-parameters, we grouped all experiments obtained in our previous paper~\cite{plaquet2024mambadiarization} by chunk size and tuned their hyper-parameters. The tuning was done on the compound training and validation sets, where we used Optuna's~\cite{akiba2019optuna} multivariate Tree-structured Parzen Estimator to find the \textit{clustering threshold} and \textit{minimum cluster size} that minimize DER. Then, for each experiment in a group with the same chunk size, we selected the hyper-parameters that achieved the best performance on the dev set and computed the median for each parameter. We used the median as the value for the hyperparameters. 
Therefore, in this paper, all experiments with the same chunk size share the same hyper-parameters (regardless of encoder, decoder or training loss).

\subsubsection{EEND architectures}

\textbf{Encoder:}
For SincNet, we used the default architecture from the pyannote package: 60 SincNet filterbanks extracted with stride 10 and kernel size 251. We initialized the SincNet with weights from the pretrained model released in \cite{plaquet23powerset}. SincNet contains 42k parameters and downsamples 16kHz 1D waveform to 59Hz 60-dimensional feature vectors.

For WavLM, we used the WavLM Base \textit{architecture} and WavLM Base or WavLM Base+ \textit{pretrained weights} depending on the experiment (the architecture is identical for both pretrained weights). WavLM Base weights are trained on 960 hours from LibriSpeech~\cite{panayotov2015librispeech}. WavLM Base+ is trained on 60 000 hours from Libri-Light~\cite{librilight}, 24 000 hours from VoxPopuli~\cite{wang2021voxpopuli}, and 10 000 hours from GigaSpeech~\cite{chen21o_interspeech}. The WavLM Base architecture contains 94M parameters (4M of those are from the convolutional feature extractor, the rest are from the 12 Transformer layers). WavLM outputs 768 features at around 50Hz, which we reduce to 256 features through a linear projection.

\textbf{Decoder:}
Our LSTM-based decoder used the architecture from \cite{plaquet23powerset}: four bidirectional LSTMs with a hidden size of 128, which amounts to 1.3M parameters. We experimented with increasing the number of layers or the hidden size to match the size of the Mamba architecture in a previous paper \cite{plaquet2024mambadiarization}, but found the LSTM-based decoder became much harder to train and led to worse performance.

For the Conformer-based model, we used the exact same architecture as in \cite{han2024leveragingselfsupervisedlearningspeaker}, four Conformer-encoder layers with four heads, kernel size 31, and position-wise feed-forward Network of size 1024. This architecture has 6.1M parameters.

For the Mamba-based decoder, we reused the architecture proposed in \cite{plaquet2024mambadiarization}: seven layers of bidirectional mamba blocks with a hidden state of size 64, while keeping other parameters default. This Mamba decoder has 8.1M parameters.

\autoref{tab:param_count} summarizes the number of parameters of every architecture we tested. Note that the chunk size does not impact the number of parameters, and the number of parameters added with the powerset loss is negligible.

\begin{table}
\centering
\caption{Size of different models in millions of parameters depending on the choice of encoder (columns) and decoder (rows).}
\begin{tabular}{|l|cc|}
    \hline
     & SincNet & WavLM\\
     \hline
    LSTM & 1.3 & 95.3 \\
    Conformer & 6.1 & 100.1\\
    Mamba & 8.1 & 102.1 \\
    \hline
\end{tabular}    
\label{tab:param_count}
\end{table}

\subsubsection{Speaker count}

\begin{table}[tb]
    \caption{Percentage of the training data with less than $N$ speakers active in one chunk (a) and $K$ speakers active in one frame (b). The configurations we used are highlighted in bold ($\geq 97\%$ coverage).}

    \subfloat[Per-chunk speaker count $N$.]{
        \centering
        \begin{tabular}{|l|llll|}
        \hline
            \# spk ($N$) & 5s & 10s & 30s & 50s \\ \hline
            $\leq$ 1 & 21.8 & 10.7 & 3.5 & 3.4 \\ 
            $\leq$ 2 & 57.3 & 37.3 & 23.1 & 32.1 \\ 
            $\leq$ 3 & 86.7 & 70.1 & 47.7 & 50.2 \\ 
            $\leq$ 4 & \textbf{99.3} & \textbf{98.2} & 95.4 & 89.8 \\ 
            $\leq$ 5 & 99.8 & 99.4 & \textbf{97.4} & 93.3 \\ 
            $\leq$ 6 & 100 & 99.9 & 99.4 & \textbf{98.3} \\ 
            $\leq$ 7 & 100 & 100 & 99.8 & 99.3 \\ 
            $\leq$ 8 & 100 & 100 & 100 & 100 \\ \hline
        \end{tabular}
        \label{tab:spk_stats_chunk}
    }\hfill
    \subfloat[Per-frame speaker count $K$.]{
        \centering
        \hspace{0.25cm}  
        \begin{tabular}{|l||c|} 
         \hline
         \makecell{\# spk ($K$)} & \% \\
         \hline
         $\leq$ 1 & 79.19 \\
         $\leq$ \textbf{2} & \textbf{97.83} \\
         $\leq$ 3 & 99.89 \\
         \hline
        \end{tabular}
        \label{tab:spk_stats_frame}
    }
    \label{tab:spk_stats}
\end{table}

The EEND decoders we used can only output a fixed number of speakers $N$ per chunk. If there are fewer than $N$ speakers in the chunk, the model will predict the superfluous speakers as inactive. When using the powerset loss, we also need to fix the maximum number of speakers in a frame $K$ (i.e. the maximum number of simultaneous speakers). This means we need to decide $N$ and $K$ before training. To do so, we computed speaker statistics at different chunk sizes on the training subsets of all our datasets and compiled them in \autoref{tab:spk_stats}. 

\autoref{tab:spk_stats_chunk} shows the proportion of training data that has $N$ or less speakers per chunk for different $N$. To compute this, we apply a sliding window over the training files and measure how many chunks contain $N$ or less speakers.

\autoref{tab:spk_stats_frame} shows the proportion of training data that contains less than $K$ at a given moment. To compute this, we discretize the reference annotation (here we used 100Hz) and count the number of frames that have $K$ or less speakers.

We chose the number of speakers for each chunk size such that at least 97\% of chunks can be correctly classified on our training data. We end up with $N=\{4,4,5,6\}$ speakers for each $W=\{5,10,30,50\}$-second chunk, respectively. We also chose the number of simultaneous speakers such that at least 97\% of the data can be correctly classified, and end up with $K=2$ simultaneous speakers for the multiclass powerset classification.

\subsection{Training settings}
\label{sec:train_settings}

The training was done on the compound training dataset, every individual dataset is sampled from with equal probability. Every model saw 72 000 samples per epoch, which represents 1000h with $W=50\text{s}$ and 100h with $W=5\text{s}$ models. 
Each minibatch contained 32 samples. The training was done for 80 epochs. We increased the learning rate (LR) linearly from $0$ to $0.002$ during the first epoch and then applied a linear cyclic LR scheduler~\cite{smith2017cyclical}, with a period of two epochs, that decays exponentially with a half-life of ten epochs. Gradients were clipped so that their norm did not exceed $1.0$. When sampling audio chunks with more than the maximum $N$ speakers supported by the network, we only kept the $N$ most active speakers.

\subsubsection{Finetuning WavLM}
We experimented with two configurations when using WavLM features. The first configuration used fixed WavLM parameters, and we only trained the decoder and final linear layers, as done in our previous study \cite{plaquet2024mambadiarization}. Because WavLM contains many parameters relative to the rest of the network, freezing it greatly speeds up training and reduces memory usage. 

In the other configuration, we unfreezed WavLM parameters so that the whole network was trained jointly. For this, we first trained a model with fixed WavLM parameters and then unfreezed the WavLM parameters and re-trained the whole system. Finetuning the WavLM allows joint optimization of the whole EEND model at the cost of more involved training.

For the configurations with a frozen WavLM, we used the settings described in \autoref{sec:train_settings}. When finetuning the WavLM, we changed the training configuration slightly. We reduced the batch size from 32 to 16 due to memory constraints. We changed the initial LR from 0.002 to 0.000125. This corresponds to an LR of $\approx 0.00025$ if we kept a batch size of 32, which is the LR after 40 epochs in our LR scheduler. We kept the same scheduler but set its half-life to 8 epochs. We trained until convergence, with an early stopping of 16 epochs without any improvement. Inspired by \cite{han2024leveragingselfsupervisedlearningspeaker}, we used label smoothing with $\alpha = 0.1$ and AutoClip \cite{seetharaman2020autoclip}.

\subsubsection{Domain adaptation}

The model training described above (with both frozen and unfrozen WavLM) was conducted on the compound training dataset, resulting in generalist diarization models. However, we used evaluation sets with different data distributions, and diarization datasets have disparate annotation standards. For these reasons, we decided to perform an additional ``domain adaptation'' using the best models on the compound development dataset. We selected one model for each decoder architecture, and for each of them obtained eight domain-adaptated models (one for each evaluation dataset).

For the adaptation, we kept the WavLM unfrozen, used a batch size of 16, fixed the LR to $0.000125$, and stopped the training process after 10 epochs with no improvement in validation DER. We also tested domain adaptation with a frozen WavLM (that has been finetuned), but no performance improvement was observed.

\subsection{Metrics}

The main metric we use in this paper is the Diarization Error Rate (DER)~\cite{nist2009evaluation} defined as \autoref{eq:der},
\begin{equation}
    \text{DER}(\hat{y}, y) = \min_\pi \frac{\text{FA}(\pi(\hat{y}), y) + \text{MISS}(\pi(\hat{y}), y) + \text{CONF}(\pi(\hat{y}), y)}{\text{TOTAL}(y)},
    \label{eq:der}
\end{equation}
where $\hat{y}$ and $y$ are predicted and reference diarization, respectively.
FA is the \textit{False Alarm}, the duration of speech that was detected but absent in the reference. MISS is the \textit{Missed Detection}, the duration of speech that was present in the reference but not predicted. CONF is the \textit{Speaker Confusion}, the duration of speech that was correctly detected but assigned to the wrong speaker. TOTAL is the total amount of speech present in the reference. 

When reporting the average DER on multiple datasets, we always report the ``macro'' average DER (compute DERs on each dataset, then average them). This is done instead of computing DER on the compound dataset since the durations of each dataset are unbalanced and DER is affected by the total speech duration in the dataset.

We also report two types of DERs: the ``Local DER'' and the ``Global DER''. We call ``Global DER'' the DER that is computed on the final pipeline prediction, after VC, $\tilde{\vect{Y}}$. This is the DER that is reported in the literature. We call ``local DER'' the DER that is computed on all local segmentations $\tilde{y}_l,$. This metric only evaluates the EEND model and disregards the vector clustering step altogether. 

Some papers \cite{tawara2024chime7, bullock2020overlapawarediar} report diarization performance with Oracle clustering, i.e., ``Oracle clustering DER'', which evaluates the pipeline prediction $\tilde{\vect{Y}}$ obtained with an oracle assignment of speaker identities instead of vector clustering. Local DER is slightly different from the Oracle clustering DER, and measures more directly the performance of the EEND model, which is the focus of our study.

\section{Results}
\label{sec:results}

In this section, we perform comparative experimental results of the different configurations of EEND-VC.
We first evaluate the impact of each factor (encoder, decoder, training loss, chunk size) on the macro performance of the validation set in \autoref{sec:expe_factors}.
We then look at the impact of WavLM finetuning in \autoref{sec:exp_wavlm_finetuning}. Finally, we compare in \autoref{sec:sota} the best configurations for each model with the SOTA on the evaluation set of each dataset.

\subsection{Evaluation of each factor}
\label{sec:expe_factors}

\begin{figure}[tb]
    \centering

    \includegraphics[width=\linewidth]{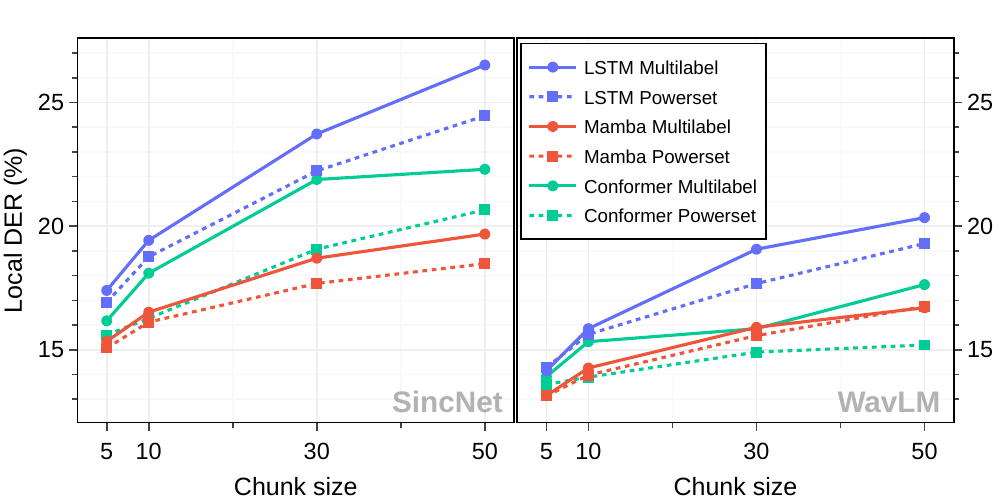}

    \begin{subfigure}[b]{0.52\textwidth}
    \caption{SincNet encoder}
    \end{subfigure}
    \begin{subfigure}[b]{0.47\textwidth}
    \caption{WavLM Base encoder}
    \end{subfigure}
    
    \caption{Local macro DER as a function of chunk size for each architecture when using (a) SincNet or (b) WavLM Base for the encoder.}
    \label{fig:allcomp_localder}
\end{figure}

\begin{figure}[t]
    \centering
    \includegraphics[width=\linewidth]{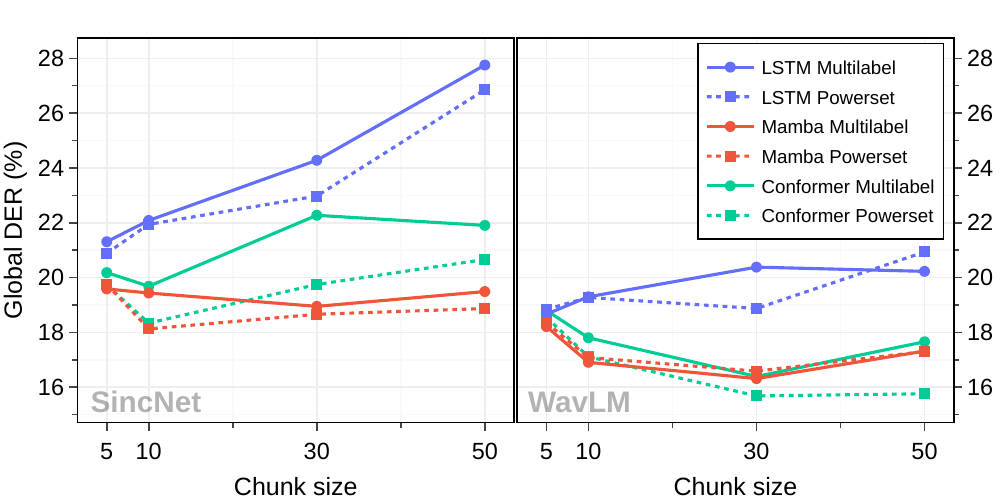}

    \begin{subfigure}[b]{0.52\textwidth}
    \caption{SincNet encoder}
    \end{subfigure}
    \begin{subfigure}[b]{0.47\textwidth}
    \caption{WavLM Base encoder}
    \end{subfigure}
    
    \caption{Global macro DER as a function of chunk size for each encoder architecture.}
    \label{fig:allcomp_globalder}
\end{figure}

In order to find the best architecture, we compare the performance of each configuration on two metrics: the ``local diarization error rate'' reported in \autoref{fig:allcomp_localder}, and the ``global diarization error rate'' reported in \autoref{fig:allcomp_globalder}.
In addition, we also include a detailed breakdown of the components of the global DER in \autoref{fig:allcomp_globalder_overlap_ml} and \autoref{fig:allcomp_globalder_overlap_ps}.
Metrics were obtained on the validation set to ensure the models selected for the SOTA comparison are not biased. The next subsections report observations for each of the four factors we studied: chunk size $W$, encoder, decoder, and training loss.

\subsubsection{Impact of chunk size}

The chunk size plays a critical role in EEND-VC. The EEND part deals with a much simplified version of the diarization problem when the chunk size is short: less context needs to be tracked and only so many distinct speakers can appear in a few seconds. This leaves the Vector Clustering part responsible for correctly identifying speakers identities and avoiding speaker confusion, but if speaker embeddings are extracted from a short context, they will be noisy and make clustering harder.
Using longer chunks reverses this dynamic, the EEND model is harder to train in this case, but if it performs well, the Vector Clustering step will be improved thanks to cleaner speaker embeddings. 
Although a critical parameter, few studies have investigated the impact of the chunk size on performance across various EEND-VC model configurations and over several datasets. 

We first analyze the influence of chunk size on our local DER results in \autoref{fig:allcomp_localder}. We observe that local DER increases with chunk size, which is expected as the task get more complex with longer chunks. Interestingly, the performance degradation varies greatly between architecture. With a SincNet encoder, which only provides features extracted from a local context, the DER degradation is severe. This degradation can also be observed with the WavLM encoder, which provides features extracted from the entire input context, but to a lesser extent. However in both cases, this degradation can be greatly mitigated using Mamba, or using Conformer with the powerset loss, which indicates these architecture better handle longer contexts than LSTM.

On the other hand, we do not observe the same straightforward trend in the global DER in \autoref{fig:allcomp_globalder}, longer chunk sizes can be beneficial. This is also to be expected, chunk size is a hyperparameter that needs to be optimized depending on the capabilities of the EEND model and the vector clustering part. Longer chunks can improve the overall diarization by reducing the speaker confusion caused by noisy embeddings, as we discuss in Section \autoref{sec:breakdown}. However, this is only the case if the increase in local DER previously observed does not outweigh the benefits of a longer chunk size for speaker embedding extractions.

From these experiments we can observe that using longer chunks can benefit the diarization pipeline when the underlying EEND architecture is powerful enough to deal with the longer context length. Using WavLM, Mamba, Conformer and the powerset loss result in models that are more resilient to long chunk size.


\subsubsection{Multilabel vs. Powerset losses}
\label{sec:ml_vs_ps}

The powerset loss has recently been favored by some papers \cite{landini2023diaper, baroudi24_interspeech, han2024leveragingselfsupervisedlearningspeaker} as it was experimentally shown to provide better performance and faster convergence at no additional cost (if the number of speakers remains relatively small). However, the original study was only conducted on a very specific configuration: 5-second LSTM-based EEND models using SincNet features.

In \autoref{fig:allcomp_localder} and \autoref{fig:allcomp_globalder}, the solid and dashed lines correspond to the results of the multilabel and powerset losses, respectively. From these figures, we can observe that using the powerset loss results in improved or comparable local and global DER in almost all cases. Its benefits are amplified with chunk length, at $W=5\text{s}$ the multilabel and powerset variants are almost equal, while they tend to have the greatest difference at $W=50\text{s}$.
The benefits seem to be greater on SincNet-based models than on WavLM-based ones. Conformer benefits from it the most, LSTM only significantly improves on SincNet-based architectures, and Mamba seems to be the least sensitive to its effects.

From these experiments, we can observe that the powerset loss outperforms the multi-label loss in most cases, and best performance for both SincNet- and WavLM-based encoders is obtained using the powerset loss.

\subsubsection{SincNet vs. WavLM encoders}
\label{sec:sincnet_vs_wavlm}

Next, we compare the SincNet and WavLM encoders.
On both local and global DER, for every architecture, the WavLM version greatly outperforms its SincNet-based counterpart. The performance gap between both increases with chunk length. WavLM makes longer chunks (i.e. 30 or 50 seconds) advantageous for EEND-VC. 

These observations align with our expectations. WavLM has been trained on much more data, and unlike SincNet's, WavLM's features are dependent on the entire input context, which makes the workload of the decoder much lighter when dealing with long context. Weaker architectures, such as LSTM, benefit the most from WavLM. In practice, this means that with short context, WavLM is slightly better than SincNet, reducing global DER by around 1\%. And with longer context, it depends on the model used, with improvements ranging from 2 to 8\% of macro average global DER. WavLM helps handling long chunks and increase overall pipeline performance, at the cost of increased inference cost. For short chunks, the lighter SincNet remains relatively competitive.

\subsubsection{LSTM vs. Mamba vs. Conformer decoders}
Finally, we compare the performance obtained with the three different decoders.
LSTM-based models largely underperform Mamba- and Conformer-based ones, both when using SincNet or WavLM Base encoders. As seen in \autoref{tab:param_count}, LSTM models contain fewer parameters than the two others, but in preliminary experiments, we confirmed that increasing the model size did not improve their performance \cite{plaquet2024mambadiarization}.
LSTMs have more limited capabilities to deal with long sequences than Conformer or Mamba. 
As a result, the gap between LSTM and other architectures increases strongly with longer chunks.

Mamba appears to be the most robust architecture overall. It works almost equally well with powerset or multilabel losses. It outperforms the other architectures with SincNet learned features, and almost matches Conformer on longer chunks with fixed WavLM features. Regardless of the encoder, it remains robust to longer chunk lengths, unlike LSTM-based architectures, despite both being RNN modules.

We achieve the best performance with the Conformer-based decoder, but only when using WavLM features and powerset loss. Since they rely on attention rather than compressing the input context into a state vector, we could expect Conformers to perform better than both LSTM and Mamba. Yet, we found the Conformer decoder do not always beat Mamba: it obtains inferior performance than Mamba when using SincNet features, and with WavLM it requires powerset loss to outperform it. 

Mamba- and Conformer-based decoders achieved close performance. While Mamba is less sensitive to the system design choices (loss, encoder, chunk size), given the right configuration, Conformer achieves the overall best performance and slightly outperforms Mamba.

\subsubsection{Breakdown of errors}
\label{sec:breakdown}
\begin{figure}[t]
    \centering
    \includegraphics[width=\linewidth]{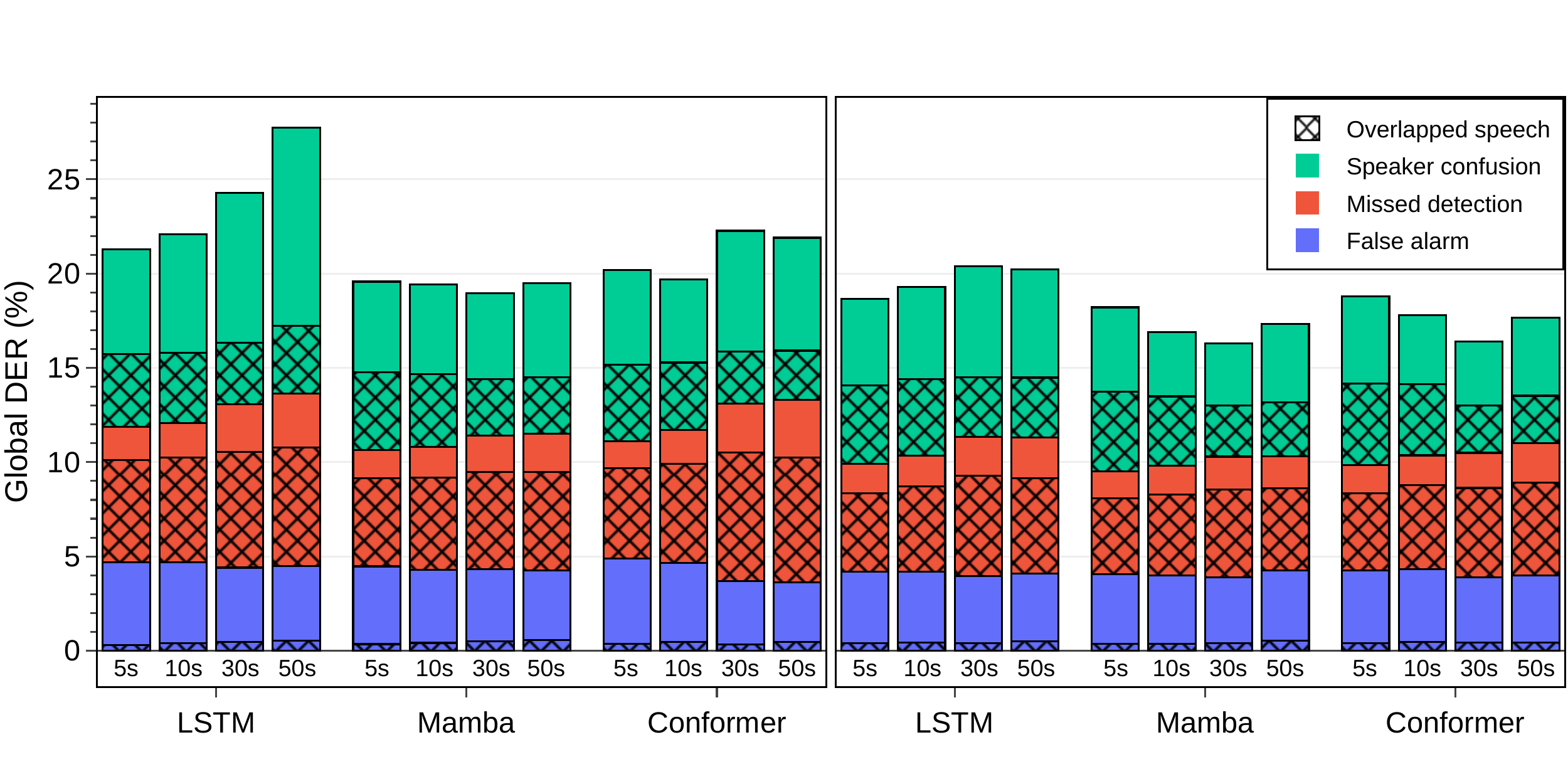}

    \begin{subfigure}[b]{0.52\textwidth}
    \caption{SincNet encoder}
    \end{subfigure}
    \begin{subfigure}[b]{0.47\textwidth}
    \caption{WavLM Base encoder}
    \end{subfigure}
    
    \caption{Detail of the DER components for models trained with the \textbf{multilabel} loss for each tested architecture with different chunk sizes and using (a) SincNet encoder and (b) WavLeM Base encoder.}
    \label{fig:allcomp_globalder_overlap_ml}
\end{figure}

\begin{figure}[t]
    \centering
    \includegraphics[width=\linewidth]{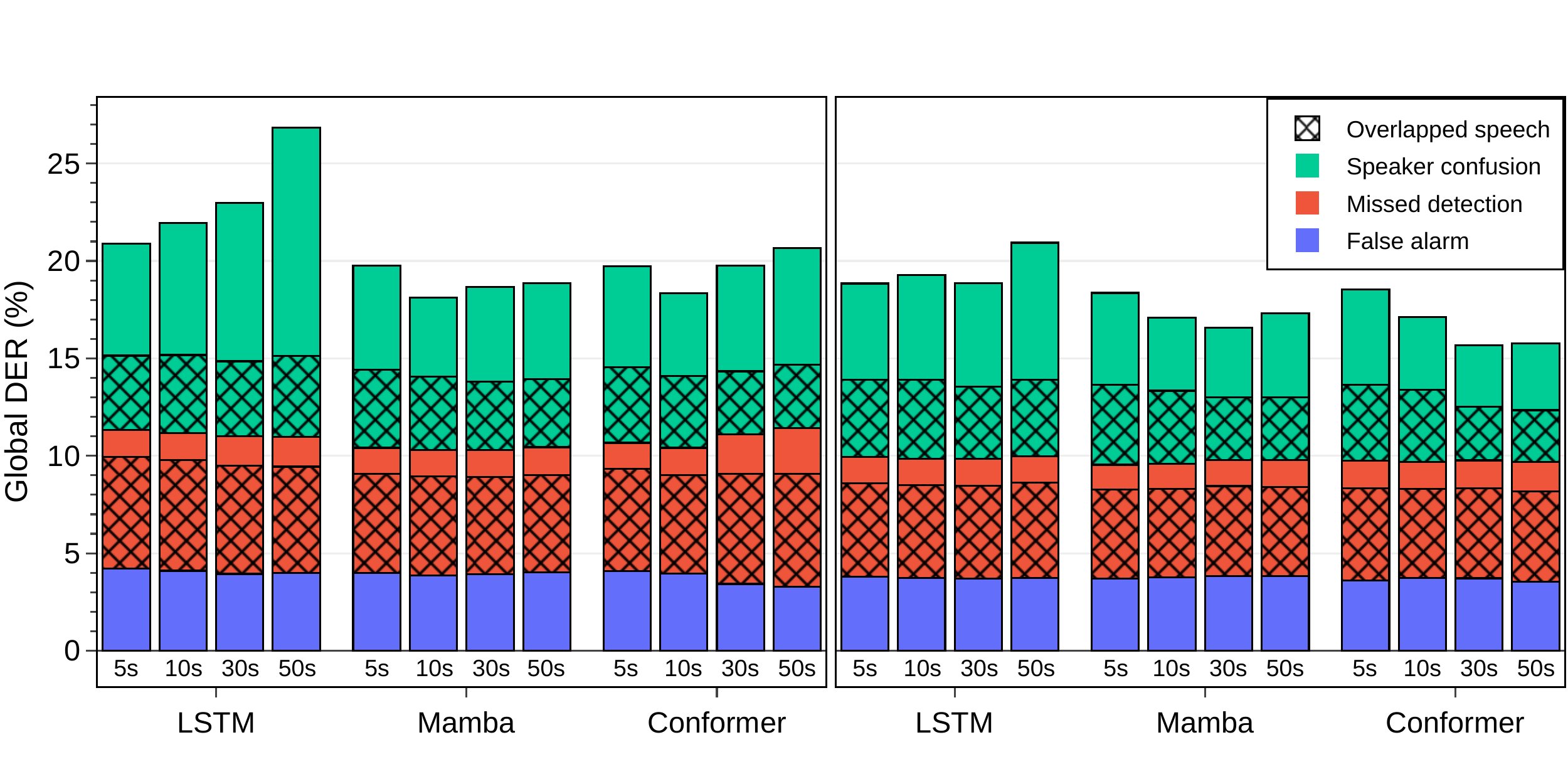}

    \begin{subfigure}[b]{0.52\textwidth}
    \caption{SincNet encoder}
    \end{subfigure}
    \begin{subfigure}[b]{0.47\textwidth}
    \caption{WavLM Base encoder}
    \end{subfigure}
    
    \caption{Detail of the DER components for models trained with the \textbf{powerset} loss for each tested architecture with different chunk sizes and using (a) SincNet encoder and (b) WavLeM Base encoder.}
    \label{fig:allcomp_globalder_overlap_ps}
\end{figure}

\autoref{fig:allcomp_globalder_overlap_ml} and \autoref{fig:allcomp_globalder_overlap_ps} plot the breakdown of error components when using the multilabel and powerset loss, respectively.
First, we can see that increasing the chunk size primarily affects the speaker confusion component. False alarm and missed detection only depend on the number of speakers predicted at each frame and should not be linked to chunk size. In practice, missed detection does degrade slightly with chunk size using the multilabel loss in \autoref{fig:allcomp_globalder_overlap_ml}. The optimal choice of $W$ minimizes speaker confusion by balancing local diarization and vector clustering quality.

Second, comparing the error components using the multilabel loss in \autoref{fig:allcomp_globalder_overlap_ml} against those obtained using the powerset loss in \autoref{fig:allcomp_globalder_overlap_ps}, it is clear that the two losses have two different distributions of error types. As described above, the chunk size mostly affects speaker confusion, which is the expected behaviour, as the false alarm and missed detection should not depend on the chunk size. We observe this exact behaviour with the powerset loss in \autoref{fig:allcomp_globalder_overlap_ps}: for a given architecture, the sum of the false alarm and missed detection remains approximately the same for all tested chunk sizes. However, this is not the case when using the multilabel loss in \autoref{fig:allcomp_globalder_overlap_ml}; missed detection increases with chunk size. 

This might indicate that models trained with the multilabel loss struggle to properly converge and learn detection of silence, speech and overlapped speech compared to those trained with powerset. The same behaviour can be observed using powerset with the \textit{(Sincnet+Conformer+Powerset)} configuration. This configuration struggles with 30 and 50 seconds chunk sizes while the same configuration using Mamba does not, despite the configuration relying on attention and slightly outclassing Mamba with a pretrained WavLM. While it is expected that training EEND models on longer chunks is harder, some architectures choices amplify this effect more than others. For example the multilabel loss, the SincNet encoder or the Conformer decoder tend to result in subpar performance with long chunks, while the powerset loss or the WavLM encoder tend to mitigate this effect.

\subsection{WavLM Finetuning}
\label{sec:exp_wavlm_finetuning}
\begin{figure}[htbp]
    \centering

    \begin{subfigure}[b]{0.49\textwidth}
        \includegraphics[width=\linewidth]{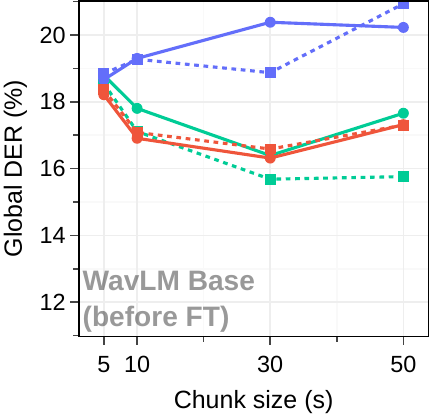}
        \caption{WavLM Base weights, no finetuning}
        \label{fig:wavlm_ft_comp1_a}
    \end{subfigure}
    \hfill
    \begin{subfigure}[b]{0.49\textwidth}
        \includegraphics[width=\linewidth]{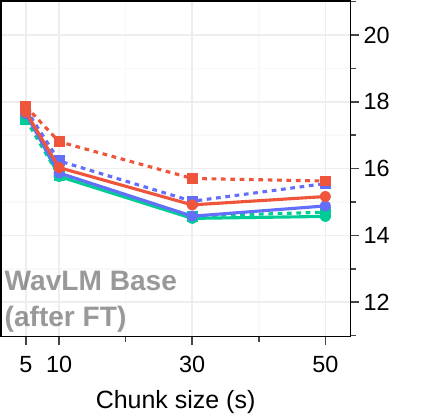}
        \caption{WavLM Base weights, finetuned}
        \label{fig:wavlm_ft_comp1_b}
    \end{subfigure}
    
    \vspace{0.5cm}
    \begin{subfigure}[b]{0.49\textwidth}
        \includegraphics[width=\linewidth]{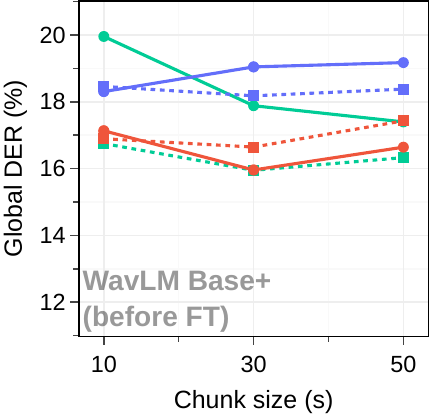}
        \caption{WavLM Base\textbf{+} weights, no finetuning}
        \label{fig:wavlm_ft_comp1_c}
    \end{subfigure}
    \hfill
    \begin{subfigure}[b]{0.49\textwidth}\
        \includegraphics[width=\linewidth]{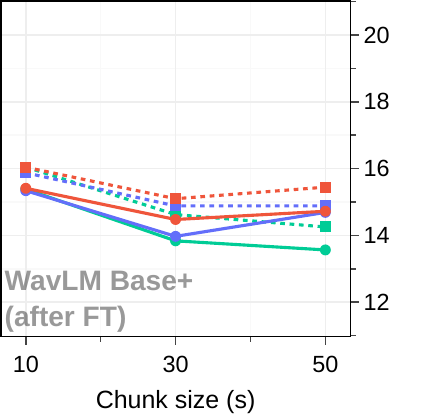}
        \caption{WavLM Base\textbf{+} weights, finetuned}
        \label{fig:wavlm_ft_comp1_d}
    \end{subfigure}
    
    \caption{Comparison of the DER obtained using pretrained feature extractors WavLM Base (a) and WavLM Base+ (c) against their respective finetuned versions (b) and (d).}
    \label{fig:wavlm_ft_comp1}
\end{figure}

\begin{figure}[tb]
    \centering
    \includegraphics[width=0.99\linewidth]{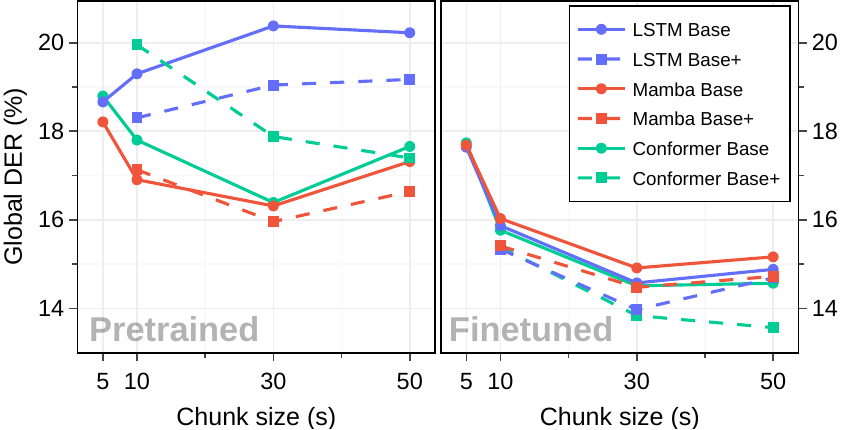}

    \begin{subfigure}[b]{0.49\textwidth}
        \caption{Trained with frozen WavLM weights}
    \end{subfigure}
    \hfill
    \begin{subfigure}[b]{0.49\textwidth}
        \caption{Finetuned with unfrozen WavLM weights}
    \end{subfigure}
    
    \caption{Global DER of WavLM-based multilabel models before (a) and after finetuning (b).}
    \label{fig:wavlm_ft_comp3}
\end{figure}

In \autoref{sec:sincnet_vs_wavlm}, we only compared SincNet to WavLM Base, and ignored WavLM Base+ for clarity and conciseness. In this section, we dive into the potential brought by allowing pretrained WavLM Base and Base+ to be finetuned through speaker diarization training (as opposed to the previous section where WavLM weights were frozen). We display the local DER of all models with WavLM Base and WavLM Base+, before and after finetuning WavLM in \autoref{fig:wavlm_ft_comp1}, and global DER before and after finetuning WavLM in \autoref{fig:wavlm_ft_comp3}. Like in the previous sections, we report the macro average DER on the validation sets.

As seen in \autoref{fig:wavlm_ft_comp1}, finetuning the WavLM encoder has a drastic positive impact on the DER, all model configurations greatly benefit from it. We can observe in \autoref{fig:wavlm_ft_comp1} that unlike previous experiments, the multilabel variants are better than their powerset counterparts after finetuning WavLM. Using a powerset makes training converge faster but at a much worse optimum. Since the powerset loss is clearly inferior in this case, we focus on the multilabel variant of finetuned WavLM-based models for the rest of the paper.

\autoref{fig:wavlm_ft_comp1_b} and \autoref{fig:wavlm_ft_comp1_d} show that all configurations are notably closer in performance after finetuning WavLM. WavLM's multiple MHSA layers may suffice to perform diarization, as they are an order of magnitude larger than our proposed decoders and have access to the uncompressed input context thanks to self-attention. While Mamba achieves competitive performance with pretrained WavLMs, it performs slightly worse than other models after finetuning. Surprisingly, LSTM performs better than Mamba and almost equals Conformer's performance, except with $W=50\text{s}$ where Conformer still has the edge.

Finetuned WavLM Base+ brings a small, but consistent improvement in performance over finetuned WavLM Base, which suggests that self-supervised models trained on more diverse data could perform even better. In the following sections, when discussing finetuned WavLM, we only focus on the WavLM Base+ variant.

\subsection{SOTA comparison}
\label{sec:sota}

\begin{table*}[t]
    \caption{DER comparison to December 2024 SOTA. (Parentheses) indicates a 0.25s collar was used for evaluation. \derbg{Bold} indicates the best-known DER (we also include systems less than 2\% worse relatively). ML and PS indicate the loss used in training, MultiLabel, or PowerSet.}
    \label{tab:sota}

    \setlength\extrarowheight{0.06cm}
    \centering
    
    \resizebox{\linewidth}{!}{%
    \begin{tabular}{|l |l l|c|c|c|c|c|c|c||>{\columncolor[gray]{0.9}}c||c|}
    \hline
        Features & System & \vthml{Parameters} & \vthml{AISHELL-4\\\textit{channel 1}} & \vthml{AliMeeting\\ \textit{far}} & \vthml{AMI\\ \textit{(channel~1)}} & \vthml{MSDWild \textit{(Few)}} &  \vth{RAMC} & \vthml{NOTSOFAR-1\\\textit{(channel 1)}} & \vthml{VoxConverse\\ \textit{v0.3}} &  \vthml{In-domain\\macro average} & \vth{DIHARD III}\\
    \hline
    \multirowcell{3}{SincNet} & (1) LSTM 5s PS & & 12.8 & 24.7 & 23.1 & 29.3 \cder{22.6} & 14.1 & 28.3 & 11.7 \cder{8.9} & 20.6 & 29.5 \nlt
     & (2) Mamba 10s PS & & 11.9 & 21.6 & 21.1 & 25.5 \cder{19.0} & 13.0 & 26.6 & 10.7 \cder{8.0} & 18.6 & 27.9 \nlt
     & (3) Conformer 10s PS & & 12.1 & 22.0 & 21.4 & 25.0 \cder{18.5} & 12.4 & 26.3 & 10.5 \cder{7.8} & 18.5 & 27.0 \nlt
    \hline
    \multirowcell{3}{WavLM\\Base} & (4) LSTM 5s ML & & 11.9 & 22.2 & 21.8 & 25.7 \cder{19.3} & 12.0 & 24.3 & 10.4 \cder{7.8} & 18.3 & 26.5 \nlt
     & (5) Mamba 30s ML & & 10.9 & 17.5 & 19.3 & 20.9 \cder{14.5} & 11.9 & 24.9 & 10.6 \cder{8.0} & 16.6 & 25.0 \nlt
     & (6) Conformer 30s PS & & 9.9 & 19.1 & 18.8 & 18.8 \cder{12.6} & 11.6 & 25.7 & 10.3 \cder{7.8} & 16.3 & 23.5 \nlt
    \hline
    \multirowcell{3}{WavLM\\Base+} & (7) LSTM 10s ML & & 11.6 & 20.9 & 20.6 & 23.0 \cder{16.4} & 11.4 & 23.7 & 9.7 \cder{7.2} & 17.3 & 25.5 \nlt
    & (8) Mamba 30s ML & & 10.8 & 18.0 & 19.1 & 20.0 \cder{13.7} & 11.1 & 25.0 & 10.3 \cder{7.7} & 16.3 & 23.5  \nlt
    & (9) Conformer 30s PS & & 10.8 & 18.6 & 19.5 & 18.6 \cder{12.4} & \derbg{10.9} & 24.3 & 10.2 \cder{7.6} & 16.1 & 23.0 \nlt
    \hline
    \multirowcell{4}{Finetuned\\WavLM\\Base+} & (10) LSTM 30s ML & & \derbg{10.2} & 13.8 & 16.5 & \derbg{17.8} \cder{11.5} & \derbg{10.9} & 20.9 & 9.8 \cder{7.4} & \derbg{14.3} & 22.2 \nlt
     & (11) Mamba 30s ML & & 11.1 & 13.7 & 16.5 & \derbg{17.7} \cder{11.5} & \derbg{10.7} & 21.9 & 10.0 \cder{7.5} & 14.5 & 22.1 \nlt
     & (12) Conformer 50s ML & & \derbg{10.3} & \derbg{12.5} & 16.3 & 18.6 \cder{12.4} & \derbg{10.9} & \derbg{19.8} & 10.6 \cder{8.0} & \derbg{14.1} & 21.3 \nlt
     & (13) \variant 30s ML & & 11.0 & 14.8 & 17.3 & 17.3 \cder{11.2} & 10.9 & 20.6 & 10.0 \cder{7.5} & 14.6 & 21.6 \nlt
    \hline
    
    \hline
    & State-of-the-art & & \makecell{10.6\\ \cite{baroudi24_interspeech}} & \makecell{13.2\\ \cite{harkonen24_eendm2f}} & \makecell{\derbg{15.4}\\ \cite{han2024leveragingselfsupervisedlearningspeaker}} & \parbox{1.5cm}{\centering \margintxt{19.6}{10.0} \\ \cite{baroudi24_interspeech}} & \makecell{11.1 \\ \cite{harkonen24_eendm2f}} & ----- & \parbox{1.5cm}{\centering \margintxt{-----}{\derbg{4.0}} \\ \cite{baroudi2023pyannote}} & \diagbox[height=2\line,width=1cm]{}{} & \makecell{\derbg{15.1} \\ \cite{cheng2024seqtoseq}} \\
    \hline
    \end{tabular}}
\end{table*}

\autoref{tab:sota} shows the global DER on the evaluation sets of the seven datasets included in the compound training data, the macro DER on these sets, and the DER on DIHARD III, which has not been seen during training. We compare our models' different configurations with SOTA. We report numbers on each tested encoder: SincNet, WavLM Base, WavLM Base+, and fine-tuned WavLM Base+. For each of the encoders, we show results with LSTM, Mamba, and Conformer, using the chunk length and training loss that achieved the best macro average global DER on the dev set for that configuration.

For SincNet-based models, we can see that the LSTM model (system 1) obtains roughly similar results to the ones reported in \cite{plaquet23powerset}, and that Mamba (system 2) and Conformer (system 3) variants greatly outperform it on all datasets. The best Mamba (system 2) and Conformer (system 3) obtain very similar performance when using SincNet features.

Using WavLM Base (systems 4--6) or WavLM Base+ (systems 7--9) instead of SincNet results in a 16\% relative improvement in DER. It also enables usage of longer context, $W=30\text{s}$ is the optimal chunk size for Mamba (systems 5, 8) and Conformer (systems 6, 9). In this case, Mamba- and Conformer-based models largely outperform the LSTM-based ones (systems 4, 7). WavLM Base+ obtains slightly better results than regular WavLM Base. With WavLM, Mamba, and Conformer also achieve comparable performance. With regular WavLM Base, Conformer may have a very slight edge and beats Mamba on 5 datasets (Mamba beats Conformer only on 2 datasets), but the difference is small with WavLM Base+.

We finally showcase the result on the finetuned WavLM Base+ (10--12), which provides another significant boost in DER with an 11\% relative improvement from models based on a frozen WavLM Base+. All three variants obtain similar performance, Conformer has a slightly better average DER but does not improve performance on all datasets. Since most of the diarization processing can be done with the WavLM module, even the LSTM-based model (10) manages to reach results competitive with SOTA. Of the three, Mamba achieves slightly worse macro DER, which comes from a worse DER on AISHELL-4 and NOTSOFAR-1, but still achieves similar results on the other datasets, and even achieves a new SOTA on MSDWild and RAMC.

The best overall model is the Conformer 50s Multilabel with finetuned WavLM Base+ (12). It outperforms the state-of-the-art results without any domain adaptation on all datasets except AMI, MSDWild, and DIHARD III. We selected this configuration because it has the best average global DER on validation sets, but we also include its $W=30\text{s}$ variant (13) to compare with LSTM (10) and Mamba (11), which perform best at 30s. In this case, at an equal chunk length, the Conformer-based model tends to perform slightly worse, meaning the larger chunk size seems to be responsible for the superiority of the system (12). 

Moreover, we can observe that on MSDWild and VoxConverse, the 50s Conformer is worse than the 30s LSTM and Mamba, but the 30s Conformer manages to equal or beat them. These results show that one chunk size does not fit all, as these two datasets are still too challenging for a longer chunk size to be advantageous, even with our best EEND configuration.

\subsection{Domain adaptation}

\begin{figure}[tb]
    \centering
    \includegraphics[width=1.0\linewidth]{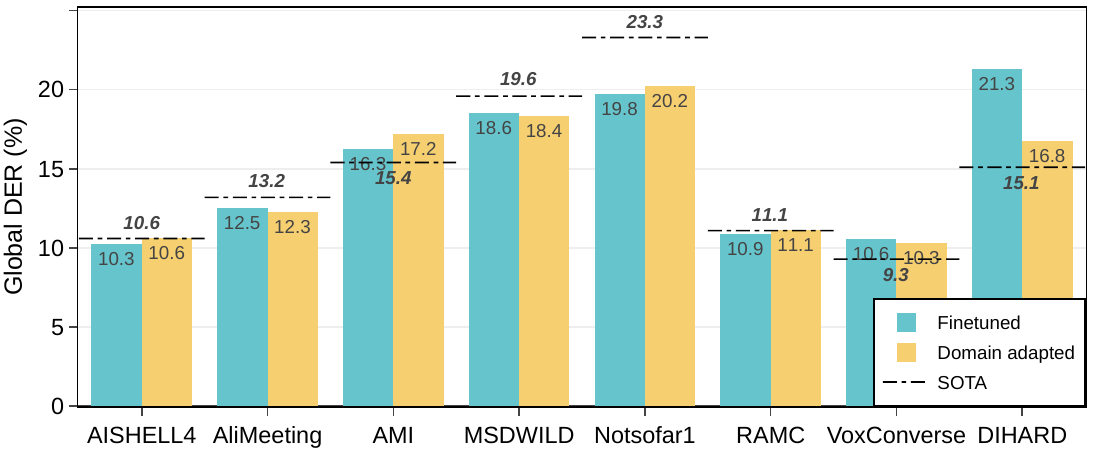}
    \caption{Comparison among the model WavLM Base+ Conformer 50s ML pretrained on the compound set, the same model after domain adaptation, and the SOTA.}
    \label{fig:domainadaptation}
\end{figure}

\begin{table}[tb]
    \setlength\extrarowheight{0.06cm}
    \caption{Performance of the best models based on a finetuned WavLM Base+, before and after domain adaptation.}
    \centering
    
    \resizebox{\linewidth}{!}{
    \begin{tabular}{|l |l |c|c|c|c|c|c|c||>{\columncolor[gray]{0.9}}c||c|}
        \hline
        Features & System & \vthml{AISHELL-4\\\textit{channel 1}} & \vthml{AliMeeting\\ \textit{far}} & \vthml{AMI\\ \textit{(channel~1)}} & \vthml{MSDWild \textit{(Few)}} &  \vth{RAMC} & \vthml{NOTSOFAR-1\\\textit{(channel 1)}} & \vthml{VoxConverse\\ \textit{v0.3}} &  \vthml{In-domain\\macro average} & \vth{DIHARD III}\\
        \hline
        \multirowcell{6}{Finetuned\\WavLM\\Base+} & (10) LSTM 30s ML & 10.2 & 13.8 & 16.5 & 17.8 & 10.9 & 20.9 & 9.8 & 14.3 & 22.2 \nlt
        & \subvariant Domain adapted & 10.0 & 13.5 & 16.1 & 17.8 & 11.0 & 20.7 & 9.7 & 14.8 & 18.2 \nlt \cline{2-11}
        & (11) Mamba 30s ML & 11.1 & 13.7 & 16.5 & 17.7 & 10.7 & 21.9 & 10.0 & 14.5 & 22.1 \nlt
        &  \subvariant Domain adapted & 11.0 & 14.0 & 16.5 & 17.7 & 10.8 & 22.0 & 9.8 & 15.1 & 18.9 \nlt \cline{2-11}
        & (12) Conformer 50s ML & 10.3 & 12.5 & 16.3 & 18.6 & 10.9 & 19.8 & 10.6 & 14.1 & 21.3 \nlt
        & \subvariant Domain adapted & 10.6 & 12.3 & 17.2 & 18.4 & 11.1 & 20.2 & 10.3 & 14.9 & 16.8 \nlt
        \hline
    \end{tabular}}
    \label{tab:domain_adaptation}
\end{table}

After obtaining EEND models with finetuned WavLMs, we do domain adaptation on each individual domain. The comparison of the best model before and after domain adaptation is shown in \autoref{fig:domainadaptation}.

This additional step does not show any consistent improvement in performance. Three of the seven datasets seen during training are slightly improved from it, and four are slightly degraded. We only observed large improvement on DIHARD, which was never seen during training. The result of domain adaptation for other decoders is also shown in \autoref{tab:domain_adaptation}, but they follow the same trend: the performance only improved on DIHARD III.

The models rely on WavLM Base+, which was trained on order of magnitudes more data than available for domain adaptation, which might explain the very marginal changes in performance.

\section{Discussion}
\label{sec:discussion}

We summarize below the key findings of our extensive evaluation.

\subsection{Lightweight models}

Models that use the SincNet-based encoder are largely outclassed by those relying on WavLM. Although their performance is inferior, they offer a much lighter alternative, which can prove useful in memory-bound scenarios or real time processing on limited devices. In this case, the safest option appears to be SincNet + Mamba + Powerset with 10s chunks. SincNet + Conformer + Powerset reaches similar results with 10s chunks, but it seems to be less resilient to longer chunk sizes.

\subsection{SOTA models}

On the other hand, if the only concern is DER, WavLM is a far superior choice. In particular, finetuning the pretrained WavLM is key to reaching SOTA results. Going from SincNet- to WavLM-based model improves DER as much as going from WavLM- to finetuned WavLM-based model. 

Our best-performing model uses a Finetuned WavLM Base+, Conformer, and Multilabel output with 50s chunks. 

\subsection{Impact of the powerset output}

Using a powerset loss instead of a multilabel one helps model performance in the vast majority of our experiments.
Our analysis suggests that powerset appears to have the most positive effect when the model struggles to learn the task.

One major evidence of this is chunk size: the longer the input chunk (which means a much harder task), the more positive impact the powerset tends to have on local DER. We also found Conformer decoders with the multilabel loss underperform: they are beaten by Mamba despite Mamba not using attention. But when using powerset loss, Conformer decoders do equal or outperform Mamba. This suggests that Conformers do not reach their full potential when trained with multilabel but do with powerset.

On the contrary, when finetuning WavLM (which brings us to SOTA results), we observe that using a powerset output results in worse performance than multilabel. In this case it seems that powerset hinders performance, and suggests that the powerset loss is easier to learn but leads to slightly suboptimal results.

We observed in \autoref{sec:breakdown} that multilabel and powerset result in two different distributions of error types, indicating that the two losses have different strengths and weaknesses. In particular, with the powerset loss the sum of false alarm and missed detection remains constant regardless of the context size, but not with the multilabel loss.

One possible explanation to these differences is that using the multiclass powerset loss makes the \textit{segmentation} part of the speaker diarization task more straightforward to learn (there are distinct classes for each possible speaker count), but the \textit{speaker identification} part more limited than with a multilabel output (the relation between classes containing the same speaker have to be deduced by the model).

\subsection{Chunk size}

We found chunk size to be a crucial hyperparameter to make EEND-VC pipelines reach better performance. It is necessary to take full advantage of more powerful EEND models that can better handle long context and distinct speaker identities. We expect future EEND-VC architectures to use increasingly long chunk sizes when possible.

In this paper, we share the same chunk size during both training and inference, but future works could explore usage of dynamic chunk size. Since the optimal chunk size for a given model vary from dataset to dataset, it might be possible to further optimize pipeline performance by training the EEND model with variable chunk length and automatically estimate the optimal length to use during inference depending on the data.

\section{Conclusion}
\label{sec:conclusion}

We analyzed the impact of multiple architecture choices of the EEND model on the performance of an EEND-VC. We found that the most impactful factor is the encoder, a finetuned WavLM Base architecture is enough to reduce global DER by around 25\% relatively, compared to a simpler SincNet-based encoder. Furthermore, when finetuned on the diarization task, WavLM greatly reduces the impact of the decoder, which in our case is an order of magnitude smaller than the WavLM.

The next most impactful factor is the choice of chunk size. Small chunk sizes are useful with difficult datasets and weaker EEND models as they put the burden of speaker identification on the clustering part of the pipeline. However, longer chunk sizes can theoretically yield much better results, but they require a more powerful EEND model to do so, otherwise they might only degrade performance.

The choice of decoder is also important, Mamba appears to be the most robust decoder, it performs well regardless of the chosen architecture. Nevertheless, it remains inferior to Conformer, which has less consistent results, but outperform Mamba by a significant margin with the right configuration.

For training loss, we find that using the multiclass powerset training loss is generally better and helps boost model performance, especially when using Conformer, LSTM or SincNet. Compared to multilabel, the multiclass powerset loss seems to help with false alarm and missed detection but increases speaker confusion. It is not strictly superior to the traditional multilabel loss, as it can underperform when WavLM-based architectures are finetuned with the WavLM unfrozen. Powerset loss might bottleneck model performance with powerful enough models, in which case multilabel is preferable.


\bibliographystyle{elsarticle-num} 
\bibliography{refs}

\end{document}